%% file: webworldvecolab.tex
\documentclass{article}
\usepackage{amssymb,alife9,pstricks}
\title{Ecolab, Webworld and self-organisation}
\author{Russell K. Standish\\School of Mathematics, University of New
  South Wales\\
r.standish@unsw.edu.au, http://parallel.hpc.unsw.edu.au/rks\vspace*{-2cm}}

\newcommand{\bbeta}{\mbox{\boldmath{$\beta$}}}   

\newcommand{\EcoLab}{{\sffamily\slshape
    \mbox{\raisebox{.5ex}{Eco}\hspace{-.4em}{\makebox[.5em]{L}ab}}}}

\newcommand{\htmladdnormallinkfoot}[2]{#1\footnote{#2}}
\def\citeyear(#1)#2{(#1)\nocite{#2}}

\newcommand{\integers}{{\Bbb Z}}

\ifx\PSTloaded\undefined
\def\PSTloaded{t}
\psset{arrowsize=.01 3.2 1.4 .3}
\psset{dotsize=.07}
\catcode`@=11

\newpsobject{PST@Border}{psline}{linewidth=.0015,linestyle=solid}
\newpsobject{PST@Axes}{psline}{linewidth=.0015,linestyle=dotted,dotsep=.004}
\newpsobject{PST@Solid}{psline}{linewidth=.0015,linestyle=solid}
\newpsobject{PST@Dashed}{psline}{linewidth=.0015,linestyle=dashed,dash=.01 .01}
\newpsobject{PST@Dotted}{psline}{linewidth=.0025,linestyle=dotted,dotsep=.008}
\newpsobject{PST@LongDash}{psline}{linewidth=.0015,linestyle=dashed,dash=.02 .01}
\newpsobject{PST@Diamond}{psdots}{linewidth=.001,linestyle=solid,dotstyle=square,dotangle=45}
\newpsobject{PST@Filldiamond}{psdots}{linewidth=.001,linestyle=solid,dotstyle=square*,dotangle=45}
\newpsobject{PST@Cross}{psdots}{linewidth=.001,linestyle=solid,dotstyle=+,dotangle=45}
\newpsobject{PST@Plus}{psdots}{linewidth=.001,linestyle=solid,dotstyle=+}
\newpsobject{PST@Square}{psdots}{linewidth=.001,linestyle=solid,dotstyle=square}
\newpsobject{PST@Circle}{psdots}{linewidth=.001,linestyle=solid,dotstyle=o}
\newpsobject{PST@Triangle}{psdots}{linewidth=.001,linestyle=solid,dotstyle=triangle}
\newpsobject{PST@Pentagon}{psdots}{linewidth=.001,linestyle=solid,dotstyle=pentagon}
\newpsobject{PST@Fillsquare}{psdots}{linewidth=.001,linestyle=solid,dotstyle=square*}
\newpsobject{PST@Fillcircle}{psdots}{linewidth=.001,linestyle=solid,dotstyle=*}
\newpsobject{PST@Filltriangle}{psdots}{linewidth=.001,linestyle=solid,dotstyle=triangle*}
\newpsobject{PST@Fillpentagon}{psdots}{linewidth=.001,linestyle=solid,dotstyle=pentagon*}
\newpsobject{PST@Arrow}{psline}{linewidth=.001,linestyle=solid}
\catcode`@=12

\fi

\begin{document}
\maketitle

\begin{abstract}
Ecolab and Webworld are both models of evolution produced by adding
evolution to ecological equations. They differ primarily in the form
of the ecological equations. Both models are self-organised to a state
where extinctions balance speciations. However, Ecolab shows evidence
of this self-organised state being critical, whereas Webworld does
not. This paper examines the self-organised states of these two models
and suggest the likely cause of the difference. Also the lifetime
distribution for a mean field version of Ecolab is computed, showing
that the fat tail of the distribution is due to coevolutionary
adaption of the species.
\end{abstract}

\vspace{-.5cm}
\section{Introduction}

In models of evolving ecologies, a ``drip feed'' of mutated species
are added to a simulation of ecological dynamics. As new species are
incorporated into the ecology, they create new links in the food web,
perturbing the system dynamics. When enough links are added, feedback
loops will form, and the simulated ecology will suffer a mass
extinction. Over time, the system {\em self organises} to a state where the
introduction of new species will be balanced by extinctions, and the
system diversity fluctuates around some mean value.

But what is this state that the system self organises to? The first
suggestion was a {\em critical} state \cite{Bak-Sneppen93}, characterised
by long range influences of a species extinction over others in the
food web. The original model of Bak and Sneppen used to illustrate
this idea is no more than a cartoon. The interactions between species
in this model had no relation to biological interactions. The
first attempt to use some real biologically inspired dynamics was
probably Ecolab \cite{Standish94}, which employed the well known
Lotka-Volterra equations, for which a quite a bit of theoretical
information is available. This model clearly self organises to a state
where speciation is balanced by extinction of
average \cite{Standish98a}, although a variation of the model
(incorporating a mechanism of specialisation) produces unbounded
growth in diversity (speciation exceeding
extinction) \cite{Standish02b}.

So is this state a critical state? One problem is that criticality in
self-organised systems is only achieved in the limit of zero driving
rate --- in this case zero mutation rate. Sole {\em et
al.} \cite{Sole-etal02} prefer the term {\em self-organised
instability}. Whilst I am sympathetic to this notion, I would also
like to point out that {\em stability} is very precise term in
dynamical systems theory, referring to the behaviour of the linearised
system around an equilibrium point. Unstable ecosystems do not have to
fall apart --- the classical Lotka-Volterra \cite{Maynard-Smith74}
limit cycle is a case in point. Rather the notion of an ecosystem
persisting in time without falling apart is captured by {\em
permanence}, for which a few modest results are known for
Lotka-Volterra systems \cite{Law-Blackford92}. So perhaps
self-organised impermanence would be a more accurate description.

Self organised critical systems are characterised by a power law
distribution of extinction avalanches, and also a power law
distribution of lifetimes. Traditionally, the presence of power law
signatures in a self-organising system is taken as evidence of
self-organised criticality. Newman \citeyear(1997){Newman97b} developed another
toy evolutionary model that exhibited power law spectra, with
neither self-organisation nor criticality in sight. However, when
the artificial constant diversity restriction is lifted in the obvious way,
self-organisation reappears \cite{Standish98a}, and the model can also
be understood as a mean field approximation of coevolutionary system
that potentially admits critical behaviour.

Ecolab demonstrates power law spectra of lifetimes \cite{Standish98a},
with an exponent of -1. However, it has proven very difficult to
measure the distribution of extinctions, as extinction avalanches
overlap in Ecolab due to the finite rate of speciation. Conversely,
studies of a similar model called Webworld claim an absence of any
power law signatures \cite{Drossel-etal01}. I have implemented the
Webworld model using the \EcoLab{} \cite{Ecolab-tech-report} simulation
system. I was similarly unable to see evidence of power law
signatures, and propose a possible explanation. 

In this paper, I show that the Fourier transform of the diversity time
series is related to the lifetime distribution. Furthermore, in the
limiting case of infinitesimal speciation, this transform is the
distribution of extinction avalanches (extinction frequency).

\section{Ecolab model}\label{ecolab-section}

We start with a generalised form of the Lotka-Volterra equation 
\begin{equation}\label{lotka-volterra}
\dot{n_i} = r_in_i - n_i\sum_j\beta_{ij}n_j.
\end{equation}
Here $n_i$ is the population of species $i$, $r_i$ is the difference
between reproduction and death and $\beta_{ij}$ is the interaction
between species $i$ and $j$.

Periodically, each species $i$ generates a number of mutant species,
proportional to $n_ir_i\mu_i$, where $\mu_i$ is the mutation rate for
species $i$. For each mutant species, the parameters $r_i$, $\beta_{ij}$, and $\mu_i$ are
mutated from the parent species according to additive or
multiplicative processes --- the exact details aren't important here,
but are described in  \cite{Standish94}.

One crucial property that is preserved by the mutation operator is
{\em boundedness} \cite{Standish98b}. Boundedness ensures that
population sizes in eq (\ref{lotka-volterra}) can never exceed a
particular limit.

It turns out that a necessary condition for permanence in eq
(\ref{lotka-volterra}) is that the matrix \bbeta{} has positive
determinant \cite{Law-Blackford92}. The determinant can be written as a
sum
\begin{equation}
\det|\bbeta| = \sum_{{\bf p}\in\mathrm{perm}(1\ldots,n)} (-1)^{s(p)}\beta_{1p_1}\beta_{2p_2}\cdots\beta_{np_n}
\end{equation}
where $\mathrm{perm}(1\ldots,n)$ is the set of permutations of the
numbers $1\ldots,n$ and $s(p)$ is the number of swaps involved in the permutation.

All diagonal terms of \bbeta{} must be positive to ensure boundedness
of eq (\ref{lotka-volterra}).

Now consider permutations with one swap ($i\rightarrow j, j\rightarrow
i$. If the terms $\beta_{ij}$ and $\beta_{ji}$ are of opposite sign
(predator-prey case), then the contribution to the determinant is
positive. However, if the terms have the same sign, (eg +ve, the
mutual competition case, an increase in $n_i$ causes $n_j$ to
decrease, which reduces competition on $n_i$, reinforcing the original
change) then it describes a positive feedback loop between species $i$
and $j$. 

Likewise, it can be seen that the term
$T = (-1)^{s(p)}\beta_{1p_1}\beta_{2p_2}\cdots\beta_{np_n}$ describes an
$s(p)$ feedback loop through the ecosystem, which is a negative
feedback loop if $T>0$, and a positive feedback loop if $T<0$. The
necessary condition for permanence can be interpreted as saying that
negative feedback loops must dominate over positive feedback loops for
the ecosystem to be permanent.

As species are added to the system through speciation, new links are
added to the foodweb at random. The chance of a positive feedback loop
forming increases dramatically as the foodweb approaches its
percolation threshold \cite{Green-Klomp99}. Once this happens, an extinction
avalanche is almost certain. The twin pressures of speciation and
extinction through impermanence oppose each other leading to a state
where the food web lies on its percolation threshold. Newth {\em et
al.} \citeyear(2002){Newth-etal02} examined the scaling structure of the Ecolab
model, and observed the critical behaviour here. This is the strongest
evidence yet that Ecolab self-organises to a critical state.

\begin{figure}
\psset{unit=.5}
\hspace{-1cm}{\tiny\input{ecohisto.tex}}
\vspace{-.5cm}
\caption{Lifetime distributions for different values of the maximum
  mutation rate $\mu$ in Ecolab. Histograms have been scaled so that the peaks
  are comparable in size.}
\label{ecolifetimes}
\vspace{-.5cm}
\end{figure}
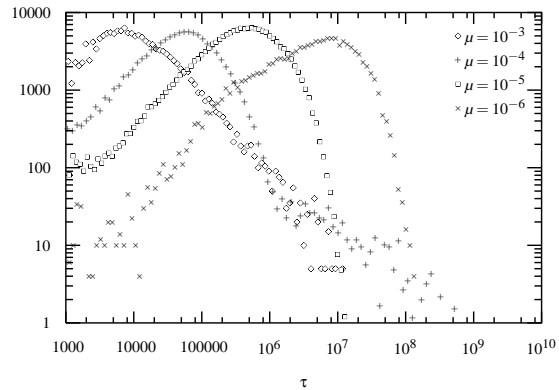

Plots of the lifetime distribution for several different values of the
maximum mutation rate (mutation rates in Ecolab are allowed to vary,
but can never exceed the maximum value) are shown in figure
\ref{ecolifetimes}. These can be compared with other published data,
such as  \cite{Standish98a}. At higher mutation rates, the distributions
exhibit a power law tail with exponent $-1$. As the mutation rate is
turned down, the power law tail disappears, leaving a lognormal
distribution. It is unclear whether the power law has disappeared
altogether, or whether with the collection of more data it will be
resolved out of the noise at the base of the graph.

\section{Relationship between diversity time series and lifetime distribution}

When a species becomes extinct, it may trigger secondary extinctions
in other species, in a chain of extinctions known as an {\em
  extinction avalanche}. In the Bak-Sneppen model, these avalanches
follow a power law distribution in avalanche size with exponent
$-1$. However, it only becomes meaningful to discuss avalanche size in
the limit of infinitesimal mutation rate, as otherwise the extinction
avalanches overlap each other. In the Ecolab case, speciation occurs
continuously, as do the resulting extinctions. More interesting is to
discuss the frequency spectrum of extinctions, obtained by Fourier
transforming the extinction time series. Diversity (number of species
in the ecosystem at any point in time) is simply the difference
between the speciation and extinction time series --- in the
infinitesimal speciation limit, the diversity spectrum is identical to
the extinction spectrum.

The diversity time series can be written as a sum over speciation
events $s_j$ and associated lifetimes $\tau_j$:
\begin{equation}
D(t) = \sum_j\Theta(t-s_j) - \Theta(t-s_j-\tau_j),
\end{equation}
where \vspace{-.5cm}
\begin{displaymath}
\Theta(x) = \left\{
\begin{array}{ll}
0 & x<0\\
\frac12 & x=0\\
1 & x>0
\end{array}
\right.
\end{displaymath}
is the usual Heaviside step
function. Taking the Fourier transform of this (and ignoring constant
factors):
\begin{equation}
\tilde{D}(k) = \sum_j \exp(iks_j) \frac{1-exp(ik\tau_j)}{ik}.
\end{equation}
Now we assume that speciation events occur every timestep (ie
$s_j=j$), and that the lifetimes $\tau_j$ are sampled from a normalised
lifetime distribution $T(\tau)$. Integrating over this distribution
yields:
\begin{equation}
\tilde{D}(k) = \sum_j\exp(ikj)\frac{1-\tilde{T}(k)}{ik} = \frac{1-\tilde{T}(k)}{(1-e^{ik})ik}.
\end{equation}

\begin{figure}[t]
\psset{unit=.5}
\hspace{-1cm}{\tiny\input{webdivFFT.tex}}
\vspace{-.5cm}
\caption{Fourier transform of a typical diversity time series in
  Webworld showing hyperbolic behaviour.}
\label{WebDFFT}
\end{figure}
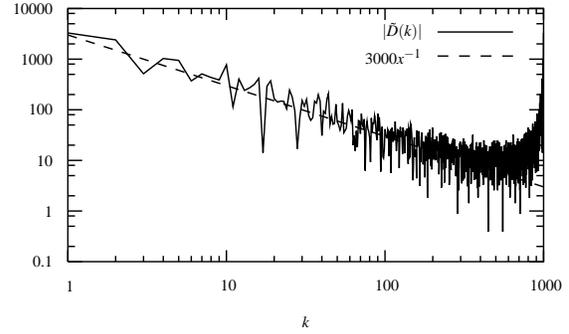

Now, if $\lim_{\tau\rightarrow0}T(\tau)<\infty$, then
$\tilde{T}(k)\rightarrow0$ as $k\rightarrow\infty$. So we can predict
that asymptotically,
\begin{equation}
|\tilde{D}(k)| \sim k^{-1} \;\;\mathrm{as}\;\;
 k\rightarrow\infty
\end{equation}
As $k\rightarrow0$, $1-e^{ik}\sim-ik$;
$1-\tilde{T}(k)\sim\langle\tau\rangle ik$, where $\langle\tau\rangle$
is the mean of $T(\tau)$. Even though a power law lifetime
distribution would lead to an infinite $\langle\tau\rangle$, and any
experiment, there is an upper cutoff to the lifetimes observed, which
would reflect a finite $\langle\tau\rangle$. Therefore also
$\tilde{D}(k)\sim k^{-1}$ as $k\rightarrow0$.  Figures \ref{EcoDFFT}
and \ref{WebDFFT} show $\tilde{D}(k)$ for typical Ecolab and Webworld
runs respectively, and both data sets demonstrate this hyperbolic
law. The power law observed in the time series spectra is
uninteresting, as it is a general feature of all such stochastic
processes. Adami \citeyear(1998){Adami98a} makes a similar point in
his book --- that power laws in the time series spectra are necessary,
but not sufficient for self-organised criticality.

\begin{figure}[t]
\psset{unit=.5}
\hspace{-1cm}{\tiny\input{ecodivFFT.tex}}
\vspace{-.5cm}
\caption{Fourier transform of a typical diversity time series in
  Ecolab showing hyperbolic behaviour.}
\label{EcoDFFT}
\end{figure}
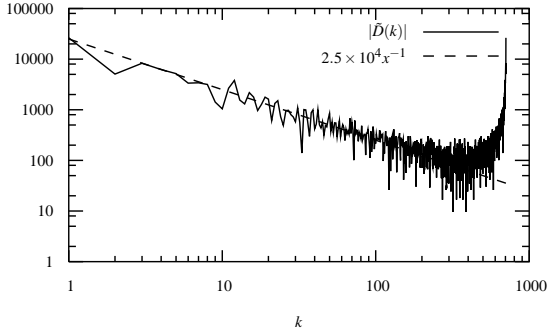

\section{Webworld}

The Webworld model was introduced by Caldarelli {\em et
  al.} \citeyear(1998){Caldarelli-etal98}, with some modifications described in
  Drossel {\em al.} \citeyear(2001){Drossel-etal01}. The model implemented here
  is taken verbatim from the latter paper, so I will give only a brief
  synopsis of the model here. The source code is available as part of
  the
  \htmladdnormallinkfoot{\EcoLab}{http://parallel.hpc.unsw.edu.au/ecolab}
  software suite.

Webworld has a population dynamics which is a generalisation of the
Lotka Volterra dynamics (eq \ref{lotka-volterra}) used in Ecolab:
\begin{equation}\label{functional-response}
\dot{n}_i=-n_i+\lambda n_i\sum_jg_{ij}(t)-\sum_jn_jg_{ji}(t).
\end{equation}
This equation is called a {\em functional response} equation.
$\lambda$ is a model parameter called {\em ecological efficiency}, and
usually taken to be $\lambda=0.1$. $n_0=R/\lambda$ is a special species, called
the {\em environment}. By choosing
$g_{ij}(t)n_i-g_{ji}(t)n_j=\beta_{ij}n_in_j, \;\;\forall i,j>0$ and
$g_{i0}=1$, equation (\ref{lotka-volterra}) is recovered. However,
unlike Ecolab, Webworld tracks resources, and so $n_i$ is perhaps
better interpreted as the amount of biomass represented by species $i$
than a population size.

In Webworld, the {\em functional response} term $g_{ij}$ is given by 
\begin{equation}\label{gdef}
g_{ij}(t)=\frac{S_{ij}f_{ij}(t)n_j(t)}{bn_j(t)+\sum_k\alpha_{ki}S_{kj}f_{kj}(t)n_k(t)}
\end{equation}
where the {\em efforts} $f_{ij}$ are given recursively:
\begin{equation}\label{fdef}
f_{ij}(t)=\frac{g_{ij}(t)}{\sum_kg_{ik}(t)}
\end{equation}
Drossel {\em et al.} \citeyear(2001){Drossel-etal01} show that allowing species
to vary the amount of effort in this way is an {\em evolutionary
  stable strategy}. The $\alpha_{ij} \leq 1$ terms above represent that
different species do not compete as strongly as members of the same
species ($\alpha_{ii}=1,\;\; \forall i$). 
\begin{equation}
\alpha_{ij}=c+(1-c)q_{ij}
\end{equation}
where $0<c<1$ is a {\em competition} parameter that strongly
influences the final steady state diversity of the model.
The precise definition of the interaction terms $S_{ij}$ and $q_{ij}$
is very interesting, but not germane to the argument here. 

In  \cite{Drossel-etal01}, the equations are evolved in time until the
ecosystem reaches equilibrium, or until a large period of time has
elapsed before another species is added to the system. This is to
mimic the ``infinitesimal speciation rate'' mentioned in relation the
self-organised criticality. In this study, a single species is added
periodically every 20 time units, which was chosen empirically as
being sufficiently rare for the ecosystem to stabilise between speciations.

\begin{figure}
\psset{unit=.5}
\hspace{-1cm}{\tiny\input{wwhisto.tex}}
\vspace{-.5cm}
\caption{Lifetime distributions for different values of $c$ for Webworld.}
\label{Weblife}
\vspace{-1cm}
\end{figure}
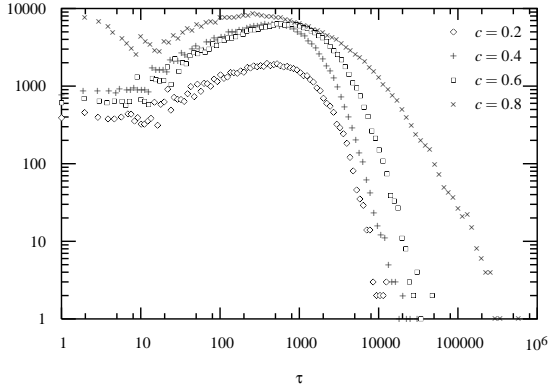

Figure \ref{WebDFFT} shows the spectrum of the diversity time series
for a run with $c=0.4$. Figure \ref{Weblife} shows a lifetime
distribution from these runs, illustrating that the best functional
form is a lognormal distribution of lifetimes. Its important to note
that lognormal distributions are often confused with power law
distributions \cite{Mitzenmacher03}, however are distinctly different
in that lognormal distributions have a mean $\langle\tau\rangle$,
whereas power law distributions do not.  However, as mentioned in the
section describing the Ecolab model, very low speciation rates were
used in this model, and it is possible that a power law tail is hidden
within the noise at tail of the graph.

Quite another explanation for the absence of critical behaviour in
Webworld comes from considering the dynamics of the effort
coefficients $f_{ij}$. These are iterated within a timestep to
determine evolutionary stable values --- this is a model of how
predators exhibiting high phenotypic plasticity might adapt to
variations in food sources. It would seem to be a less plausible model
of less complex organisms that might not be so choosy about their food
source. Whatever the biological realism of this process, the $f_{ij}$
coefficients describe the effective foodweb, which tends to be quite
sparse and without many loops \cite{Quince-etal02}. Could it be that
this process prevents the percolation threshold of the foodweb from
being reached?

\section{Mean Field Ecolab model}

In  \cite{Standish98a}, I introduced a mean field\footnote{Some people
  use {\em mean field} to mean {\em panmictic}. The Ecolab model
  described in previously is already panmictic. By mean
  field, I mean that each species experiences a stochastic force that
  is the average of the interspecific interactions in the full model.}
  version of the Ecolab model (which I dubbed Ecolab\verb+--+). That model is
  a simple multiplicative process, which is related by logarithms to
  the standard isotropic 1D random walk process. The lifetime
  distribution is known as the {\em first-passage time} distribution
  in this subject, and is known to exhibit a $\tau^{-3/2}$
  tail \cite{Redner01}. Note that this is different from, but still
  compatible with, the upper bound of $\tau^{-1}$ derived in
   \cite{Standish98a}.

However, this model has a lognormal limiting distribution of
population sizes $n$ of the form:
\begin{equation}
p(n,t) = \frac1{n\sqrt{t}}\exp(-\frac{(\ln n-rt)^2}{4t})
\end{equation}
This distribution does not satisfy boundedness.

In order to introduce boundedness, we need to reintroduce the
quadratic term into the mean field model:
\begin{equation}\label{meanfree}
\dot{n}=n(r-\beta n + \gamma),
\end{equation}
where $\gamma$ is an uncorrelated stochastic variable, with zero mean.

Taking logarithms $\xi=\ln n$ and applying the Ito transformation
formula \cite[p.  372, e.g.]{Karlin-Taylor81}, eq (\ref{meanfree}) can
be written as a Langevin equation:
\begin{equation}
\dot{\xi}   = (r-\frac12-\beta e^\xi + \gamma).
\end{equation}
The extra term of $\frac12$ comes from the effect of change of
variables on the stochastic term $\gamma$. Langevin equations can be converted into a
Fokker-Planck equation describing the probability distribution
$w(x,t)$ that $\xi$ has the value $x$ at time $t$ \cite{Risken84}:
\begin{equation}\label{Fokker-Planck}
\frac{\partial w}{\partial t} = \frac{\partial^2 w}{\partial t^2} - 
\frac{\partial}{\partial t}(r - \frac12-\beta e^x)w
\end{equation}

Taking the Laplace transform of equation (\ref{Fokker-Planck}) yields
a second order homogeneous ordinary linear differential equation:
\begin{equation}\label{trans-Fokker-Planck}
\frac{\partial}{\partial x}\left(w'(x,s) - (r-\frac12-\beta e^x)w\right) + sw = 0.
\end{equation}

The full time dependent equation doesn't appear to be amenable to
analytic treatment, however the time independent equation ($s=0$) can
be reduced to a 1st order ODE. Let
\begin{eqnarray}\label{w_0}
y(x) &=& \exp\left((r-\frac12)x-\beta e^x\right) \\
y'(x) &=& \left(r-\frac12-\beta e^x\right)y(x) = g(x)y(x),
\end{eqnarray}
and write $w_0(x)=w(x,0)=y(x)v(x)$. Substitute this into equation
(\ref{trans-Fokker-Planck}), and one obtains:
\begin{eqnarray}
\frac{d}{dx}(yv') &=& 0 \nonumber \\
v(x) &=& A_0\int^x\frac{dx'}{y(x')} + A_1\nonumber\\
     &=& A_0\left(-\beta\right)^{r-\frac12}\Gamma(\frac12-r,-\beta e^x) + A_1
\end{eqnarray}
where $\Gamma(a,x)$ is an incomplete gamma
function \cite[6.5.3]{Abramowitz-Stegun65}, and $A_j$ are constants of
integration.  Substituting this into the expression for $p(n,0)$
yields:
\begin{eqnarray}
\lefteqn{p(n,0) = \frac1n w(\ln n,0) =}&&\nonumber\\
&& n^{r-\frac32}e^{-\beta
  n}\left(A_1+A_0\left(-\beta\right)^{r-\frac12}\Gamma(\frac12-r,-\beta n)\right)
\end{eqnarray}
From the series
$\Gamma(a,x)\sim\Gamma(a)-\frac{x^a}{a}+\frac{x^{a+1}}{a+1}\cdots$  \cite[8.354]{Gradsteyn-Ryzhik80},
one can see:
\begin{eqnarray}
p(n,0)&\sim& n^{r-\frac32}e^{-\beta n}
\left(A_0\left(-\beta\right)^{r-\frac12}\Gamma(\frac12-r)+A_1\right) +\nonumber\\
&&\frac{A_0}{n(r-\frac12)}e^{-\beta n} 
\end{eqnarray}
which is normalisable if and only if $r>\frac12$ and $A_0=0$. We may
therefore set $w_0(x)=y(x)$.

The asymptotic behaviour at large times translates into the the small
$s$ regime. We can compute $w_1(x)=\left.\frac{\partial w(x,s)}{\partial
  s}\right|_{s=0}$ by differentiating eq (\ref{trans-Fokker-Planck})
with respect to $s$.
\begin{equation}
\frac{d}{dx}(w_1'-(r-\frac12-\beta e^x)w_1) + w_0 = 0
\end{equation}
to which the solution is:
\begin{equation}\label{w1}
w_1(x)=w_0(x)\left(A_1+\int^x\frac1{w_0}(x')\int^{x'}w_0(x'')dx''\right).
\end{equation}

The innermost integral can be evaluated, the answer
being \cite[6.5.2]{Abramowitz-Stegun65}
\begin{equation}
\int^xw_0(x')dx' = \beta^{-r+\frac12}\gamma(r-\frac12,\beta e^x)
\end{equation}
with $\gamma$ being another of the incomplete gamma functions. This can be
represented as a series \cite{Gradsteyn-Ryzhik80}:
\begin{equation}
\int^xw_0(x')dx' =
\beta^{-r-\frac12}\sum_{n=0}^\infty \frac{(-1)^n(\beta e^x)^{r-\frac12+n}}{n!(r-\frac12+n)}
\end{equation}
Performing the integral on each term of the series yields:
\begin{eqnarray}
\lefteqn{w_1(x)=w_0(x)\times}\\
&&\left(A_1+\exp(\beta
e^x)\sum_{n=1}^\infty\sum_{j=0}^{n-1}\frac{(-1)^j\beta^je^{jx}}{j!n(r-\frac12+n)}+E_1(\beta e^x)\right)\nonumber
\end{eqnarray}
where $E_1(x)$ is the exponential integral  \cite[5.1.1]{Abramowitz-Stegun65}.

Interestingly, for the special case $r-\frac12\in\integers^+$, the result can
be expressed as a finite series. It might seem that $r$ can be chosen
to any value by scaling the time dimension without loss of generality,
however that is not the case, as the timescale is already set by the
variance of the stochastic term $\gamma$ in
eq. (\ref{meanfree}). 

Considering the special case $r=\frac32$, and making use of the identity
$\gamma(1,x)=(1-e^{-x})$ \cite[8.352]{Gradsteyn-Ryzhik80}, we have:
\begin{equation}\label{w1-r}
w_1(x)=w_0(x)(A_1 - \beta^{-1}e^{-x}-\Gamma(-1,\beta e^x))
\end{equation}


To compute the asymptotic form of the lifetime distribution, we make
use of {\em first passage theory} \cite{Redner01}. The first passage
probability $F(n,t|n_0)$ of the population having value $n$, given a
starting value $n_0$ at $t=0$ is related to $p(n,t|n_0)$:
\begin{equation}
p(n,t) = \delta(t)\delta(n-n_0) + \int_0^t
F(n,t-t'|n_0)p(n_0,t|n_0)dt'
\end{equation}
Taking the Laplace transform, and rearranging gives us
\begin{equation}
\tilde{F}(n,s) = p(n,s) / p(n_0,s) \; \; n_0\neq n
\end{equation}
(as we're not interested in the $n=n_0$ case). For concreteness, let
$n_0=10$ and $n=1$, as is taken in the case of Ecolab experiments
computing the lifetime distribution (Fig. \ref{ecolifetimes}). The
asymptotic form can be computed directly from $F(n,s)$:
\begin{equation}
\tilde{F}(n,1/\tau) = \int^\infty e^{-\tau t}F(n,t)dt 
\sim \int^\tau F(n,t)dt  
\end{equation}
\vspace{-5mm}
\begin{eqnarray}
\lefteqn{F(n,\tau)  \sim  \frac{\partial}{\partial\tau} \tilde{F}(n,1/\tau)}\\
    &&= -\frac{p(n_0,0)\left.\frac{\partial
            p(n,s)}{\partial s}\right|_{s=0}-p(n,0)\left.\frac{\partial
            p(n_0,s)}{\partial
            s}\right|_{s=0}}{\tau^2p(n_0,0)^2}\nonumber
\end{eqnarray}
Unless the numerator vanishes, the long time tail will obey a
$\tau^{-2}$ power law. Substituting equations (\ref{w_0}) and
(\ref{w1-r}) yields for the case $r=\frac32$:
\begin{equation}
F(n,\tau)\sim \frac{e^{2\beta n_0}}{\tau^{-2}}
\left(\frac1{\beta n_0}-\frac1{\beta n} +
\Gamma(-1,n_0)-\Gamma(-1,n)\right) 
\end{equation}
Since $\Gamma(-1,n)>\Gamma(-1,n_0)$ for $n<n_0$, this derivative term
is negative. It seems unlikely to vanish for any value of $r>0$,
however this will need to be checked numerically.
 
This result is interesting. The mean field model can be considered as
a neutral model, in the sense of the neutral shadow models proposed by
Bedau and Packard \cite{Bedau-etal98}. An observed excess of lifetimes over
the mean field case (in Fig \ref{ecolifetimes} a $\tau^{-1}$
distribution is observed) would represent coadaption by the species in
the ecosystem.

\section{Conclusion}

In this paper, I consider the question of self-organised criticality
in a couple of evolutionary ecology models (Ecolab and Webworld).
In spite of their similarity, only Ecolab appears to self-organise to
criticality, whereas Webworld's self-organised state appears to be
noncritical, in agreement with Webworld's creator's statements. 

Whilst it is possible that experiments have not been run long enough
to observe critical behaviour, a more likely explanation is organismal
plasticity in Webworld prevents long range interdependence of species
in the foodweb from building up.

A mean free approximation to the Ecolab model is solved, and the
lifetime distribution from this model is expected to have a
$\tau^{-2}$ asymptotic behaviour. The fact that the real Ecolab model
appears to have a $\tau^{-1}$ asymptotic behaviour hints at adaption
occurring within that system.

Finally, the spectral density of the diversity time series (which is
related to the distribution of extinction avalanches) is expected to
have a $1/f$ behaviour, regardless of the underlying process, so this
should not be taken as evidence of self-organised criticality.

\vspace{-.2cm}
\section*{Acknowledgements}
\vspace{-.2cm}
I would like to thank the {\em Australian Centre for Advanced
  Computing and Communications} (ac3) for computer time use in these
  simulations. I would also like to thank Ben Goldys for helpful
  comments on the manuscript.
\vspace{-.35cm}

\bibliographystyle{alife9}
\bibliography{rus}
\end{document}

%% file: ecohisto.tex
\ifx\PSTloaded\undefined
\def\PSTloaded{t}
\psset{arrowsize=.01 3.2 1.4 .3}
\psset{dotsize=.01}
\catcode`@=11

\newpsobject{PST@Border}{psline}{linewidth=.0015,linestyle=solid}
\newpsobject{PST@Axes}{psline}{linewidth=.0015,linestyle=dotted,dotsep=.004}
\newpsobject{PST@Solid}{psline}{linewidth=.0015,linestyle=solid}
\newpsobject{PST@Dashed}{psline}{linewidth=.0015,linestyle=dashed,dash=.01 .01}
\newpsobject{PST@Dotted}{psline}{linewidth=.0025,linestyle=dotted,dotsep=.008}
\newpsobject{PST@LongDash}{psline}{linewidth=.0015,linestyle=dashed,dash=.02 .01}
\newpsobject{PST@Diamond}{psdots}{linewidth=.001,linestyle=solid,dotstyle=square,dotangle=45}
\newpsobject{PST@Filldiamond}{psdots}{linewidth=.001,linestyle=solid,dotstyle=square*,dotangle=45}
\newpsobject{PST@Cross}{psdots}{linewidth=.001,linestyle=solid,dotstyle=+,dotangle=45}
\newpsobject{PST@Plus}{psdots}{linewidth=.001,linestyle=solid,dotstyle=+}
\newpsobject{PST@Square}{psdots}{linewidth=.001,linestyle=solid,dotstyle=square}
\newpsobject{PST@Circle}{psdots}{linewidth=.001,linestyle=solid,dotstyle=o}
\newpsobject{PST@Triangle}{psdots}{linewidth=.001,linestyle=solid,dotstyle=triangle}
\newpsobject{PST@Pentagon}{psdots}{linewidth=.001,linestyle=solid,dotstyle=pentagon}
\newpsobject{PST@Fillsquare}{psdots}{linewidth=.001,linestyle=solid,dotstyle=square*}
\newpsobject{PST@Fillcircle}{psdots}{linewidth=.001,linestyle=solid,dotstyle=*}
\newpsobject{PST@Filltriangle}{psdots}{linewidth=.001,linestyle=solid,dotstyle=triangle*}
\newpsobject{PST@Fillpentagon}{psdots}{linewidth=.001,linestyle=solid,dotstyle=pentagon*}
\newpsobject{PST@Arrow}{psline}{linewidth=.001,linestyle=solid}
\catcode`@=12

\fi
\psset{unit=5.0in,xunit=5.0in,yunit=3.0in}
\pspicture(0.000000,0.000000)(0.700000,0.700000)
\ifx\nofigs\undefined
\catcode`@=11

\PST@Border(0.1490,0.1260)
(0.1640,0.1260)

\PST@Border(0.6470,0.1260)
(0.6320,0.1260)

\rput[r](0.1330,0.1260){ 1}
\PST@Border(0.1490,0.1668)
(0.1565,0.1668)

\PST@Border(0.6470,0.1668)
(0.6395,0.1668)

\PST@Border(0.1490,0.1906)
(0.1565,0.1906)

\PST@Border(0.6470,0.1906)
(0.6395,0.1906)

\PST@Border(0.1490,0.2076)
(0.1565,0.2076)

\PST@Border(0.6470,0.2076)
(0.6395,0.2076)

\PST@Border(0.1490,0.2207)
(0.1565,0.2207)

\PST@Border(0.6470,0.2207)
(0.6395,0.2207)

\PST@Border(0.1490,0.2314)
(0.1565,0.2314)

\PST@Border(0.6470,0.2314)
(0.6395,0.2314)

\PST@Border(0.1490,0.2405)
(0.1565,0.2405)

\PST@Border(0.6470,0.2405)
(0.6395,0.2405)

\PST@Border(0.1490,0.2484)
(0.1565,0.2484)

\PST@Border(0.6470,0.2484)
(0.6395,0.2484)

\PST@Border(0.1490,0.2553)
(0.1565,0.2553)

\PST@Border(0.6470,0.2553)
(0.6395,0.2553)

\PST@Border(0.1490,0.2615)
(0.1640,0.2615)

\PST@Border(0.6470,0.2615)
(0.6320,0.2615)

\rput[r](0.1330,0.2615){ 10}
\PST@Border(0.1490,0.3023)
(0.1565,0.3023)

\PST@Border(0.6470,0.3023)
(0.6395,0.3023)

\PST@Border(0.1490,0.3261)
(0.1565,0.3261)

\PST@Border(0.6470,0.3261)
(0.6395,0.3261)

\PST@Border(0.1490,0.3431)
(0.1565,0.3431)

\PST@Border(0.6470,0.3431)
(0.6395,0.3431)

\PST@Border(0.1490,0.3562)
(0.1565,0.3562)

\PST@Border(0.6470,0.3562)
(0.6395,0.3562)

\PST@Border(0.1490,0.3669)
(0.1565,0.3669)

\PST@Border(0.6470,0.3669)
(0.6395,0.3669)

\PST@Border(0.1490,0.3760)
(0.1565,0.3760)

\PST@Border(0.6470,0.3760)
(0.6395,0.3760)

\PST@Border(0.1490,0.3839)
(0.1565,0.3839)

\PST@Border(0.6470,0.3839)
(0.6395,0.3839)

\PST@Border(0.1490,0.3908)
(0.1565,0.3908)

\PST@Border(0.6470,0.3908)
(0.6395,0.3908)

\PST@Border(0.1490,0.3970)
(0.1640,0.3970)

\PST@Border(0.6470,0.3970)
(0.6320,0.3970)

\rput[r](0.1330,0.3970){ 100}
\PST@Border(0.1490,0.4378)
(0.1565,0.4378)

\PST@Border(0.6470,0.4378)
(0.6395,0.4378)

\PST@Border(0.1490,0.4616)
(0.1565,0.4616)

\PST@Border(0.6470,0.4616)
(0.6395,0.4616)

\PST@Border(0.1490,0.4786)
(0.1565,0.4786)

\PST@Border(0.6470,0.4786)
(0.6395,0.4786)

\PST@Border(0.1490,0.4917)
(0.1565,0.4917)

\PST@Border(0.6470,0.4917)
(0.6395,0.4917)

\PST@Border(0.1490,0.5024)
(0.1565,0.5024)

\PST@Border(0.6470,0.5024)
(0.6395,0.5024)

\PST@Border(0.1490,0.5115)
(0.1565,0.5115)

\PST@Border(0.6470,0.5115)
(0.6395,0.5115)

\PST@Border(0.1490,0.5194)
(0.1565,0.5194)

\PST@Border(0.6470,0.5194)
(0.6395,0.5194)

\PST@Border(0.1490,0.5263)
(0.1565,0.5263)

\PST@Border(0.6470,0.5263)
(0.6395,0.5263)

\PST@Border(0.1490,0.5325)
(0.1640,0.5325)

\PST@Border(0.6470,0.5325)
(0.6320,0.5325)

\rput[r](0.1330,0.5325){ 1000}
\PST@Border(0.1490,0.5733)
(0.1565,0.5733)

\PST@Border(0.6470,0.5733)
(0.6395,0.5733)

\PST@Border(0.1490,0.5971)
(0.1565,0.5971)

\PST@Border(0.6470,0.5971)
(0.6395,0.5971)

\PST@Border(0.1490,0.6141)
(0.1565,0.6141)

\PST@Border(0.6470,0.6141)
(0.6395,0.6141)

\PST@Border(0.1490,0.6272)
(0.1565,0.6272)

\PST@Border(0.6470,0.6272)
(0.6395,0.6272)

\PST@Border(0.1490,0.6379)
(0.1565,0.6379)

\PST@Border(0.6470,0.6379)
(0.6395,0.6379)

\PST@Border(0.1490,0.6470)
(0.1565,0.6470)

\PST@Border(0.6470,0.6470)
(0.6395,0.6470)

\PST@Border(0.1490,0.6549)
(0.1565,0.6549)

\PST@Border(0.6470,0.6549)
(0.6395,0.6549)

\PST@Border(0.1490,0.6618)
(0.1565,0.6618)

\PST@Border(0.6470,0.6618)
(0.6395,0.6618)

\PST@Border(0.1490,0.6680)
(0.1640,0.6680)

\PST@Border(0.6470,0.6680)
(0.6320,0.6680)

\rput[r](0.1330,0.6680){ 10000}
\PST@Border(0.1490,0.1260)
(0.1490,0.1460)

\PST@Border(0.1490,0.6680)
(0.1490,0.6480)

\rput(0.1490,0.0840){ 1000}
\PST@Border(0.1704,0.1260)
(0.1704,0.1360)

\PST@Border(0.1704,0.6680)
(0.1704,0.6580)

\PST@Border(0.1987,0.1260)
(0.1987,0.1360)

\PST@Border(0.1987,0.6680)
(0.1987,0.6580)

\PST@Border(0.2132,0.1260)
(0.2132,0.1360)

\PST@Border(0.2132,0.6680)
(0.2132,0.6580)

\PST@Border(0.2201,0.1260)
(0.2201,0.1460)

\PST@Border(0.2201,0.6680)
(0.2201,0.6480)

\rput(0.2201,0.0840){ 10000}
\PST@Border(0.2416,0.1260)
(0.2416,0.1360)

\PST@Border(0.2416,0.6680)
(0.2416,0.6580)

\PST@Border(0.2699,0.1260)
(0.2699,0.1360)

\PST@Border(0.2699,0.6680)
(0.2699,0.6580)

\PST@Border(0.2844,0.1260)
(0.2844,0.1360)

\PST@Border(0.2844,0.6680)
(0.2844,0.6580)

\PST@Border(0.2913,0.1260)
(0.2913,0.1460)

\PST@Border(0.2913,0.6680)
(0.2913,0.6480)

\rput(0.2913,0.0840){ 100000}
\PST@Border(0.3127,0.1260)
(0.3127,0.1360)

\PST@Border(0.3127,0.6680)
(0.3127,0.6580)

\PST@Border(0.3410,0.1260)
(0.3410,0.1360)

\PST@Border(0.3410,0.6680)
(0.3410,0.6580)

\PST@Border(0.3555,0.1260)
(0.3555,0.1360)

\PST@Border(0.3555,0.6680)
(0.3555,0.6580)

\PST@Border(0.3624,0.1260)
(0.3624,0.1460)

\PST@Border(0.3624,0.6680)
(0.3624,0.6480)

\rput(0.3624,0.0840){$10^{6}$}
\PST@Border(0.3838,0.1260)
(0.3838,0.1360)

\PST@Border(0.3838,0.6680)
(0.3838,0.6580)

\PST@Border(0.4122,0.1260)
(0.4122,0.1360)

\PST@Border(0.4122,0.6680)
(0.4122,0.6580)

\PST@Border(0.4267,0.1260)
(0.4267,0.1360)

\PST@Border(0.4267,0.6680)
(0.4267,0.6580)

\PST@Border(0.4336,0.1260)
(0.4336,0.1460)

\PST@Border(0.4336,0.6680)
(0.4336,0.6480)

\rput(0.4336,0.0840){$10^{7}$}
\PST@Border(0.4550,0.1260)
(0.4550,0.1360)

\PST@Border(0.4550,0.6680)
(0.4550,0.6580)

\PST@Border(0.4833,0.1260)
(0.4833,0.1360)

\PST@Border(0.4833,0.6680)
(0.4833,0.6580)

\PST@Border(0.4978,0.1260)
(0.4978,0.1360)

\PST@Border(0.4978,0.6680)
(0.4978,0.6580)

\PST@Border(0.5047,0.1260)
(0.5047,0.1460)

\PST@Border(0.5047,0.6680)
(0.5047,0.6480)

\rput(0.5047,0.0840){$10^{8}$}
\PST@Border(0.5261,0.1260)
(0.5261,0.1360)

\PST@Border(0.5261,0.6680)
(0.5261,0.6580)

\PST@Border(0.5544,0.1260)
(0.5544,0.1360)

\PST@Border(0.5544,0.6680)
(0.5544,0.6580)

\PST@Border(0.5690,0.1260)
(0.5690,0.1360)

\PST@Border(0.5690,0.6680)
(0.5690,0.6580)

\PST@Border(0.5759,0.1260)
(0.5759,0.1460)

\PST@Border(0.5759,0.6680)
(0.5759,0.6480)

\rput(0.5759,0.0840){$10^{9}$}
\PST@Border(0.5973,0.1260)
(0.5973,0.1360)

\PST@Border(0.5973,0.6680)
(0.5973,0.6580)

\PST@Border(0.6256,0.1260)
(0.6256,0.1360)

\PST@Border(0.6256,0.6680)
(0.6256,0.6580)

\PST@Border(0.6401,0.1260)
(0.6401,0.1360)

\PST@Border(0.6401,0.6680)
(0.6401,0.6580)

\PST@Border(0.6470,0.1260)
(0.6470,0.1460)

\PST@Border(0.6470,0.6680)
(0.6470,0.6480)

\rput(0.6470,0.0840){$10^{10}$}
\PST@Border(0.1490,0.1260)
(0.6470,0.1260)
(0.6470,0.6680)
(0.1490,0.6680)
(0.1490,0.1260)

\rput(0.3980,0.0210){$\tau$}
\rput[r](0.6310,0.6270){$\mu=10^{-3}$}
\PST@Diamond(0.1515,0.5827)
\PST@Diamond(0.1552,0.5441)
\PST@Diamond(0.1589,0.5809)
\PST@Diamond(0.1625,0.5748)
\PST@Diamond(0.1662,0.5804)
\PST@Diamond(0.1699,0.6134)
\PST@Diamond(0.1736,0.5844)
\PST@Diamond(0.1773,0.6166)
\PST@Diamond(0.1809,0.6043)
\PST@Diamond(0.1846,0.6223)
\PST@Diamond(0.1883,0.6277)
\PST@Diamond(0.1920,0.6319)
\PST@Diamond(0.1957,0.6329)
\PST@Diamond(0.1994,0.6312)
\PST@Diamond(0.2030,0.6357)
\PST@Diamond(0.2067,0.6360)
\PST@Diamond(0.2104,0.6409)
\PST@Diamond(0.2141,0.6320)
\PST@Diamond(0.2178,0.6318)
\PST@Diamond(0.2214,0.6265)
\PST@Diamond(0.2251,0.6284)
\PST@Diamond(0.2288,0.6246)
\PST@Diamond(0.2325,0.6182)
\PST@Diamond(0.2362,0.6137)
\PST@Diamond(0.2399,0.6123)
\PST@Diamond(0.2435,0.6042)
\PST@Diamond(0.2472,0.6021)
\PST@Diamond(0.2509,0.6000)
\PST@Diamond(0.2546,0.5957)
\PST@Diamond(0.2583,0.5899)
\PST@Diamond(0.2619,0.5839)
\PST@Diamond(0.2656,0.5786)
\PST@Diamond(0.2693,0.5695)
\PST@Diamond(0.2730,0.5634)
\PST@Diamond(0.2767,0.5642)
\PST@Diamond(0.2804,0.5499)
\PST@Diamond(0.2840,0.5422)
\PST@Diamond(0.2877,0.5280)
\PST@Diamond(0.2914,0.5269)
\PST@Diamond(0.2951,0.5138)
\PST@Diamond(0.2988,0.5168)
\PST@Diamond(0.3024,0.5095)
\PST@Diamond(0.3061,0.4940)
\PST@Diamond(0.3098,0.4905)
\PST@Diamond(0.3135,0.4855)
\PST@Diamond(0.3172,0.4748)
\PST@Diamond(0.3209,0.4681)
\PST@Diamond(0.3245,0.4420)
\PST@Diamond(0.3282,0.4554)
\PST@Diamond(0.3319,0.4348)
\PST@Diamond(0.3356,0.4247)
\PST@Diamond(0.3393,0.4348)
\PST@Diamond(0.3429,0.4363)
\PST@Diamond(0.3466,0.4168)
\PST@Diamond(0.3503,0.3970)
\PST@Diamond(0.3540,0.4052)
\PST@Diamond(0.3577,0.3999)
\PST@Diamond(0.3614,0.3908)
\PST@Diamond(0.3650,0.3562)
\PST@Diamond(0.3687,0.3908)
\PST@Diamond(0.3724,0.3801)
\PST@Diamond(0.3761,0.3716)
\PST@Diamond(0.3798,0.3261)
\PST@Diamond(0.3834,0.3352)
\PST@Diamond(0.3871,0.3618)
\PST@Diamond(0.3908,0.3023)
\PST@Diamond(0.3945,0.3352)
\PST@Diamond(0.3982,0.2615)
\PST@Diamond(0.4018,0.3154)
\PST@Diamond(0.4055,0.2207)
\PST@Diamond(0.4092,0.3431)
\PST@Diamond(0.4129,0.3023)
\PST@Diamond(0.4166,0.2207)
\PST@Diamond(0.4203,0.2207)
\PST@Diamond(0.4239,0.2854)
\PST@Diamond(0.4276,0.2207)
\PST@Diamond(0.4313,0.2207)
\PST@Diamond(0.4387,0.2207)
\PST@Diamond(0.5595,0.6270)
\rput[r](0.6310,0.5850){$\mu=10^{-4}$}
\PST@Plus(0.1510,0.4654)
\PST@Plus(0.1559,0.4619)
\PST@Plus(0.1608,0.4726)
\PST@Plus(0.1656,0.4703)
\PST@Plus(0.1705,0.4784)
\PST@Plus(0.1754,0.4861)
\PST@Plus(0.1803,0.5031)
\PST@Plus(0.1851,0.4957)
\PST@Plus(0.1900,0.5192)
\PST@Plus(0.1949,0.5159)
\PST@Plus(0.1998,0.5354)
\PST@Plus(0.2047,0.5397)
\PST@Plus(0.2095,0.5564)
\PST@Plus(0.2144,0.5596)
\PST@Plus(0.2193,0.5668)
\PST@Plus(0.2242,0.5799)
\PST@Plus(0.2290,0.5863)
\PST@Plus(0.2339,0.5913)
\PST@Plus(0.2388,0.6026)
\PST@Plus(0.2437,0.6061)
\PST@Plus(0.2485,0.6156)
\PST@Plus(0.2534,0.6198)
\PST@Plus(0.2583,0.6252)
\PST@Plus(0.2632,0.6288)
\PST@Plus(0.2681,0.6330)
\PST@Plus(0.2729,0.6346)
\PST@Plus(0.2778,0.6342)
\PST@Plus(0.2827,0.6324)
\PST@Plus(0.2876,0.6280)
\PST@Plus(0.2924,0.6223)
\PST@Plus(0.2973,0.6138)
\PST@Plus(0.3022,0.6071)
\PST@Plus(0.3071,0.5979)
\PST@Plus(0.3119,0.5853)
\PST@Plus(0.3168,0.5729)
\PST@Plus(0.3217,0.5600)
\PST@Plus(0.3266,0.5441)
\PST@Plus(0.3315,0.5299)
\PST@Plus(0.3363,0.5116)
\PST@Plus(0.3412,0.4852)
\PST@Plus(0.3461,0.4689)
\PST@Plus(0.3510,0.4389)
\PST@Plus(0.3558,0.4139)
\PST@Plus(0.3607,0.3723)
\PST@Plus(0.3656,0.3552)
\PST@Plus(0.3705,0.3258)
\PST@Plus(0.3753,0.3423)
\PST@Plus(0.3802,0.3101)
\PST@Plus(0.3851,0.3377)
\PST@Plus(0.3900,0.2955)
\PST@Plus(0.3949,0.3134)
\PST@Plus(0.3997,0.3335)
\PST@Plus(0.4046,0.3181)
\PST@Plus(0.4095,0.3174)
\PST@Plus(0.4144,0.3101)
\PST@Plus(0.4192,0.3056)
\PST@Plus(0.4241,0.3261)
\PST@Plus(0.4290,0.2944)
\PST@Plus(0.4339,0.2827)
\PST@Plus(0.4387,0.3008)
\PST@Plus(0.4436,0.2714)
\PST@Plus(0.4485,0.2542)
\PST@Plus(0.4534,0.2706)
\PST@Plus(0.4582,0.2585)
\PST@Plus(0.4631,0.2281)
\PST@Plus(0.4680,0.2496)
\PST@Plus(0.4729,0.2738)
\PST@Plus(0.4778,0.1561)
\PST@Plus(0.4826,0.2625)
\PST@Plus(0.4875,0.2595)
\PST@Plus(0.4924,0.2187)
\PST@Plus(0.4973,0.2697)
\PST@Plus(0.5021,0.1837)
\PST@Plus(0.5070,0.1997)
\PST@Plus(0.5119,0.1351)
\PST@Plus(0.5216,0.1668)
\PST@Plus(0.5265,0.1938)
\PST@Plus(0.5314,0.2123)
\PST@Plus(0.5412,0.1715)
\PST@Plus(0.5558,0.1499)
\PST@Plus(0.5595,0.5850)
\rput[r](0.6310,0.5430){$\mu=10^{-5}$}
\PST@Square(0.1527,0.3847)
\PST@Square(0.1565,0.4180)
\PST@Square(0.1602,0.4067)
\PST@Square(0.1640,0.4032)
\PST@Square(0.1677,0.3911)
\PST@Square(0.1715,0.4154)
\PST@Square(0.1752,0.3991)
\PST@Square(0.1789,0.3939)
\PST@Square(0.1827,0.4121)
\PST@Square(0.1864,0.4047)
\PST@Square(0.1902,0.4175)
\PST@Square(0.1939,0.4223)
\PST@Square(0.1977,0.4171)
\PST@Square(0.2014,0.4313)
\PST@Square(0.2052,0.4351)
\PST@Square(0.2089,0.4405)
\PST@Square(0.2126,0.4561)
\PST@Square(0.2164,0.4567)
\PST@Square(0.2201,0.4674)
\PST@Square(0.2239,0.4779)
\PST@Square(0.2276,0.4788)
\PST@Square(0.2314,0.4871)
\PST@Square(0.2351,0.5030)
\PST@Square(0.2388,0.5035)
\PST@Square(0.2426,0.5091)
\PST@Square(0.2463,0.5147)
\PST@Square(0.2501,0.5212)
\PST@Square(0.2538,0.5284)
\PST@Square(0.2576,0.5379)
\PST@Square(0.2613,0.5422)
\PST@Square(0.2651,0.5501)
\PST@Square(0.2688,0.5590)
\PST@Square(0.2725,0.5648)
\PST@Square(0.2763,0.5688)
\PST@Square(0.2800,0.5744)
\PST@Square(0.2838,0.5836)
\PST@Square(0.2875,0.5875)
\PST@Square(0.2913,0.5944)
\PST@Square(0.2950,0.5997)
\PST@Square(0.2988,0.6057)
\PST@Square(0.3025,0.6128)
\PST@Square(0.3062,0.6158)
\PST@Square(0.3100,0.6209)
\PST@Square(0.3137,0.6260)
\PST@Square(0.3175,0.6293)
\PST@Square(0.3212,0.6327)
\PST@Square(0.3250,0.6365)
\PST@Square(0.3287,0.6374)
\PST@Square(0.3324,0.6406)
\PST@Square(0.3362,0.6395)
\PST@Square(0.3399,0.6405)
\PST@Square(0.3437,0.6413)
\PST@Square(0.3474,0.6394)
\PST@Square(0.3512,0.6386)
\PST@Square(0.3549,0.6355)
\PST@Square(0.3587,0.6334)
\PST@Square(0.3624,0.6270)
\PST@Square(0.3661,0.6227)
\PST@Square(0.3699,0.6167)
\PST@Square(0.3736,0.6082)
\PST@Square(0.3774,0.6049)
\PST@Square(0.3811,0.5962)
\PST@Square(0.3849,0.5869)
\PST@Square(0.3886,0.5746)
\PST@Square(0.3924,0.5635)
\PST@Square(0.3961,0.5497)
\PST@Square(0.3998,0.5341)
\PST@Square(0.4036,0.5192)
\PST@Square(0.4073,0.4922)
\PST@Square(0.4111,0.4754)
\PST@Square(0.4148,0.4423)
\PST@Square(0.4186,0.4158)
\PST@Square(0.4223,0.3833)
\PST@Square(0.4260,0.3538)
\PST@Square(0.4298,0.3120)
\PST@Square(0.4335,0.2454)
\PST@Square(0.4373,0.2183)
\PST@Square(0.4410,0.1367)
\PST@Square(0.5595,0.5430)
\rput[r](0.6310,0.5010){$\mu=10^{-6}$}
\PST@Cross(0.1528,0.2314)
\PST@Cross(0.1569,0.2615)
\PST@Cross(0.1610,0.3335)
\PST@Cross(0.1651,0.3299)
\PST@Cross(0.1733,0.2076)
\PST@Cross(0.1774,0.2076)
\PST@Cross(0.1815,0.2615)
\PST@Cross(0.1856,0.2722)
\PST@Cross(0.1897,0.2615)
\PST@Cross(0.1937,0.3023)
\PST@Cross(0.1978,0.3023)
\PST@Cross(0.2019,0.2615)
\PST@Cross(0.2060,0.2813)
\PST@Cross(0.2101,0.2615)
\PST@Cross(0.2142,0.3513)
\PST@Cross(0.2183,0.3079)
\PST@Cross(0.2224,0.2615)
\PST@Cross(0.2265,0.2076)
\PST@Cross(0.2306,0.3261)
\PST@Cross(0.2347,0.3629)
\PST@Cross(0.2388,0.3369)
\PST@Cross(0.2429,0.3607)
\PST@Cross(0.2470,0.3867)
\PST@Cross(0.2511,0.4026)
\PST@Cross(0.2552,0.3760)
\PST@Cross(0.2593,0.3824)
\PST@Cross(0.2634,0.3970)
\PST@Cross(0.2675,0.4216)
\PST@Cross(0.2716,0.4037)
\PST@Cross(0.2757,0.4282)
\PST@Cross(0.2798,0.4434)
\PST@Cross(0.2839,0.4721)
\PST@Cross(0.2880,0.4746)
\PST@Cross(0.2921,0.4673)
\PST@Cross(0.2962,0.4933)
\PST@Cross(0.3003,0.4919)
\PST@Cross(0.3044,0.5046)
\PST@Cross(0.3085,0.5091)
\PST@Cross(0.3126,0.5159)
\PST@Cross(0.3167,0.5348)
\PST@Cross(0.3208,0.5263)
\PST@Cross(0.3249,0.5451)
\PST@Cross(0.3290,0.5372)
\PST@Cross(0.3331,0.5471)
\PST@Cross(0.3372,0.5494)
\PST@Cross(0.3413,0.5540)
\PST@Cross(0.3454,0.5578)
\PST@Cross(0.3495,0.5610)
\PST@Cross(0.3535,0.5623)
\PST@Cross(0.3576,0.5656)
\PST@Cross(0.3617,0.5772)
\PST@Cross(0.3658,0.5790)
\PST@Cross(0.3699,0.5860)
\PST@Cross(0.3740,0.5878)
\PST@Cross(0.3781,0.5869)
\PST@Cross(0.3822,0.5878)
\PST@Cross(0.3863,0.5966)
\PST@Cross(0.3904,0.6006)
\PST@Cross(0.3945,0.6065)
\PST@Cross(0.3986,0.6067)
\PST@Cross(0.4027,0.6102)
\PST@Cross(0.4068,0.6123)
\PST@Cross(0.4109,0.6184)
\PST@Cross(0.4150,0.6184)
\PST@Cross(0.4191,0.6198)
\PST@Cross(0.4232,0.6222)
\PST@Cross(0.4273,0.6226)
\PST@Cross(0.4314,0.6227)
\PST@Cross(0.4355,0.6113)
\PST@Cross(0.4396,0.6170)
\PST@Cross(0.4437,0.6120)
\PST@Cross(0.4478,0.6077)
\PST@Cross(0.4519,0.5997)
\PST@Cross(0.4560,0.5953)
\PST@Cross(0.4601,0.5819)
\PST@Cross(0.4642,0.5736)
\PST@Cross(0.4683,0.5624)
\PST@Cross(0.4724,0.5511)
\PST@Cross(0.4765,0.5271)
\PST@Cross(0.4806,0.4886)
\PST@Cross(0.4847,0.4786)
\PST@Cross(0.4888,0.4541)
\PST@Cross(0.4929,0.4193)
\PST@Cross(0.4970,0.3585)
\PST@Cross(0.5011,0.3369)
\PST@Cross(0.5052,0.2892)
\PST@Cross(0.5093,0.2615)
\PST@Cross(0.5133,0.2076)
\PST@Cross(0.5595,0.5010)
\catcode`@=12
\fi
\endpspicture

%% file: ecodivFFT.tex
\ifx\PSTloaded\undefined
\def\PSTloaded{t}
\psset{arrowsize=.01 3.2 1.4 .3}
\psset{dotsize=.01}
\catcode`@=11

\newpsobject{PST@Border}{psline}{linewidth=.0015,linestyle=solid}
\newpsobject{PST@Axes}{psline}{linewidth=.0015,linestyle=dotted,dotsep=.004}
\newpsobject{PST@Solid}{psline}{linewidth=.0015,linestyle=solid}
\newpsobject{PST@Dashed}{psline}{linewidth=.0015,linestyle=dashed,dash=.01 .01}
\newpsobject{PST@Dotted}{psline}{linewidth=.0025,linestyle=dotted,dotsep=.008}
\newpsobject{PST@LongDash}{psline}{linewidth=.0015,linestyle=dashed,dash=.02 .01}
\newpsobject{PST@Diamond}{psdots}{linewidth=.001,linestyle=solid,dotstyle=square,dotangle=45}
\newpsobject{PST@Filldiamond}{psdots}{linewidth=.001,linestyle=solid,dotstyle=square*,dotangle=45}
\newpsobject{PST@Cross}{psdots}{linewidth=.001,linestyle=solid,dotstyle=+,dotangle=45}
\newpsobject{PST@Plus}{psdots}{linewidth=.001,linestyle=solid,dotstyle=+}
\newpsobject{PST@Square}{psdots}{linewidth=.001,linestyle=solid,dotstyle=square}
\newpsobject{PST@Circle}{psdots}{linewidth=.001,linestyle=solid,dotstyle=o}
\newpsobject{PST@Triangle}{psdots}{linewidth=.001,linestyle=solid,dotstyle=triangle}
\newpsobject{PST@Pentagon}{psdots}{linewidth=.001,linestyle=solid,dotstyle=pentagon}
\newpsobject{PST@Fillsquare}{psdots}{linewidth=.001,linestyle=solid,dotstyle=square*}
\newpsobject{PST@Fillcircle}{psdots}{linewidth=.001,linestyle=solid,dotstyle=*}
\newpsobject{PST@Filltriangle}{psdots}{linewidth=.001,linestyle=solid,dotstyle=triangle*}
\newpsobject{PST@Fillpentagon}{psdots}{linewidth=.001,linestyle=solid,dotstyle=pentagon*}
\newpsobject{PST@Arrow}{psline}{linewidth=.001,linestyle=solid}
\catcode`@=12

\fi
\psset{unit=5.0in,xunit=5.0in,yunit=3.0in}
\pspicture(0.000000,0.000000)(0.700000,0.600000)
\ifx\nofigs\undefined
\catcode`@=11

\PST@Border(0.1650,0.1260)
(0.1800,0.1260)

\PST@Border(0.6470,0.1260)
(0.6320,0.1260)

\rput[r](0.1490,0.1260){ 1}
\PST@Border(0.1650,0.1526)
(0.1725,0.1526)

\PST@Border(0.6470,0.1526)
(0.6395,0.1526)

\PST@Border(0.1650,0.1878)
(0.1725,0.1878)

\PST@Border(0.6470,0.1878)
(0.6395,0.1878)

\PST@Border(0.1650,0.2058)
(0.1725,0.2058)

\PST@Border(0.6470,0.2058)
(0.6395,0.2058)

\PST@Border(0.1650,0.2144)
(0.1800,0.2144)

\PST@Border(0.6470,0.2144)
(0.6320,0.2144)

\rput[r](0.1490,0.2144){ 10}
\PST@Border(0.1650,0.2410)
(0.1725,0.2410)

\PST@Border(0.6470,0.2410)
(0.6395,0.2410)

\PST@Border(0.1650,0.2762)
(0.1725,0.2762)

\PST@Border(0.6470,0.2762)
(0.6395,0.2762)

\PST@Border(0.1650,0.2942)
(0.1725,0.2942)

\PST@Border(0.6470,0.2942)
(0.6395,0.2942)

\PST@Border(0.1650,0.3028)
(0.1800,0.3028)

\PST@Border(0.6470,0.3028)
(0.6320,0.3028)

\rput[r](0.1490,0.3028){ 100}
\PST@Border(0.1650,0.3294)
(0.1725,0.3294)

\PST@Border(0.6470,0.3294)
(0.6395,0.3294)

\PST@Border(0.1650,0.3646)
(0.1725,0.3646)

\PST@Border(0.6470,0.3646)
(0.6395,0.3646)

\PST@Border(0.1650,0.3826)
(0.1725,0.3826)

\PST@Border(0.6470,0.3826)
(0.6395,0.3826)

\PST@Border(0.1650,0.3912)
(0.1800,0.3912)

\PST@Border(0.6470,0.3912)
(0.6320,0.3912)

\rput[r](0.1490,0.3912){ 1000}
\PST@Border(0.1650,0.4178)
(0.1725,0.4178)

\PST@Border(0.6470,0.4178)
(0.6395,0.4178)

\PST@Border(0.1650,0.4530)
(0.1725,0.4530)

\PST@Border(0.6470,0.4530)
(0.6395,0.4530)

\PST@Border(0.1650,0.4710)
(0.1725,0.4710)

\PST@Border(0.6470,0.4710)
(0.6395,0.4710)

\PST@Border(0.1650,0.4796)
(0.1800,0.4796)

\PST@Border(0.6470,0.4796)
(0.6320,0.4796)

\rput[r](0.1490,0.4796){ 10000}
\PST@Border(0.1650,0.5062)
(0.1725,0.5062)

\PST@Border(0.6470,0.5062)
(0.6395,0.5062)

\PST@Border(0.1650,0.5414)
(0.1725,0.5414)

\PST@Border(0.6470,0.5414)
(0.6395,0.5414)

\PST@Border(0.1650,0.5594)
(0.1725,0.5594)

\PST@Border(0.6470,0.5594)
(0.6395,0.5594)

\PST@Border(0.1650,0.5680)
(0.1800,0.5680)

\PST@Border(0.6470,0.5680)
(0.6320,0.5680)

\rput[r](0.1490,0.5680){ 100000}
\PST@Border(0.1650,0.1260)
(0.1650,0.1460)

\PST@Border(0.1650,0.5680)
(0.1650,0.5480)

\rput(0.1650,0.0840){ 1}
\PST@Border(0.2134,0.1260)
(0.2134,0.1360)

\PST@Border(0.2134,0.5680)
(0.2134,0.5580)

\PST@Border(0.2417,0.1260)
(0.2417,0.1360)

\PST@Border(0.2417,0.5680)
(0.2417,0.5580)

\PST@Border(0.2617,0.1260)
(0.2617,0.1360)

\PST@Border(0.2617,0.5680)
(0.2617,0.5580)

\PST@Border(0.2773,0.1260)
(0.2773,0.1360)

\PST@Border(0.2773,0.5680)
(0.2773,0.5580)

\PST@Border(0.2900,0.1260)
(0.2900,0.1360)

\PST@Border(0.2900,0.5680)
(0.2900,0.5580)

\PST@Border(0.3008,0.1260)
(0.3008,0.1360)

\PST@Border(0.3008,0.5680)
(0.3008,0.5580)

\PST@Border(0.3101,0.1260)
(0.3101,0.1360)

\PST@Border(0.3101,0.5680)
(0.3101,0.5580)

\PST@Border(0.3183,0.1260)
(0.3183,0.1360)

\PST@Border(0.3183,0.5680)
(0.3183,0.5580)

\PST@Border(0.3257,0.1260)
(0.3257,0.1460)

\PST@Border(0.3257,0.5680)
(0.3257,0.5480)

\rput(0.3257,0.0840){ 10}
\PST@Border(0.3740,0.1260)
(0.3740,0.1360)

\PST@Border(0.3740,0.5680)
(0.3740,0.5580)

\PST@Border(0.4023,0.1260)
(0.4023,0.1360)

\PST@Border(0.4023,0.5680)
(0.4023,0.5580)

\PST@Border(0.4224,0.1260)
(0.4224,0.1360)

\PST@Border(0.4224,0.5680)
(0.4224,0.5580)

\PST@Border(0.4380,0.1260)
(0.4380,0.1360)

\PST@Border(0.4380,0.5680)
(0.4380,0.5580)

\PST@Border(0.4507,0.1260)
(0.4507,0.1360)

\PST@Border(0.4507,0.5680)
(0.4507,0.5580)

\PST@Border(0.4614,0.1260)
(0.4614,0.1360)

\PST@Border(0.4614,0.5680)
(0.4614,0.5580)

\PST@Border(0.4708,0.1260)
(0.4708,0.1360)

\PST@Border(0.4708,0.5680)
(0.4708,0.5580)

\PST@Border(0.4790,0.1260)
(0.4790,0.1360)

\PST@Border(0.4790,0.5680)
(0.4790,0.5580)

\PST@Border(0.4863,0.1260)
(0.4863,0.1460)

\PST@Border(0.4863,0.5680)
(0.4863,0.5480)

\rput(0.4863,0.0840){ 100}
\PST@Border(0.5347,0.1260)
(0.5347,0.1360)

\PST@Border(0.5347,0.5680)
(0.5347,0.5580)

\PST@Border(0.5630,0.1260)
(0.5630,0.1360)

\PST@Border(0.5630,0.5680)
(0.5630,0.5580)

\PST@Border(0.5831,0.1260)
(0.5831,0.1360)

\PST@Border(0.5831,0.5680)
(0.5831,0.5580)

\PST@Border(0.5986,0.1260)
(0.5986,0.1360)

\PST@Border(0.5986,0.5680)
(0.5986,0.5580)

\PST@Border(0.6114,0.1260)
(0.6114,0.1360)

\PST@Border(0.6114,0.5680)
(0.6114,0.5580)

\PST@Border(0.6221,0.1260)
(0.6221,0.1360)

\PST@Border(0.6221,0.5680)
(0.6221,0.5580)

\PST@Border(0.6314,0.1260)
(0.6314,0.1360)

\PST@Border(0.6314,0.5680)
(0.6314,0.5580)

\PST@Border(0.6396,0.1260)
(0.6396,0.1360)

\PST@Border(0.6396,0.5680)
(0.6396,0.5580)

\PST@Border(0.6470,0.1260)
(0.6470,0.1460)

\PST@Border(0.6470,0.5680)
(0.6470,0.5480)

\rput(0.6470,0.0840){ 1000}
\PST@Border(0.1650,0.1260)
(0.6470,0.1260)
(0.6470,0.5680)
(0.1650,0.5680)
(0.1650,0.1260)

\rput(0.4060,0.0210){$k$}
\rput[r](0.5200,0.5270){$|\tilde{D}(k)|$}
\PST@Solid(0.5360,0.5270)
(0.6150,0.5270)

\PST@Solid(0.1650,0.5168)
(0.1650,0.5168)
(0.2134,0.4535)
(0.2417,0.4724)
(0.2617,0.4614)
(0.2773,0.4543)
(0.2900,0.4377)
(0.3008,0.4383)
(0.3101,0.4354)
(0.3183,0.4046)
(0.3257,0.3928)
(0.3323,0.4288)
(0.3384,0.4424)
(0.3440,0.4082)
(0.3491,0.4014)
(0.3540,0.4211)
(0.3585,0.4133)
(0.3627,0.3899)
(0.3667,0.4005)
(0.3705,0.4132)
(0.3740,0.4098)
(0.3774,0.3723)
(0.3807,0.4023)
(0.3838,0.4070)
(0.3868,0.3790)
(0.3896,0.3640)
(0.3923,0.3898)
(0.3950,0.3902)
(0.3975,0.3827)
(0.4000,0.3787)
(0.4023,0.3704)
(0.4046,0.3946)
(0.4068,0.3693)
(0.4090,0.3156)
(0.4111,0.3915)
(0.4131,0.3910)
(0.4150,0.3716)
(0.4170,0.3448)
(0.4188,0.3860)
(0.4206,0.3909)
(0.4224,0.3579)
(0.4241,0.3663)
(0.4258,0.3912)
(0.4274,0.3648)
(0.4290,0.3594)
(0.4306,0.3704)
(0.4321,0.3465)
(0.4337,0.3637)
(0.4351,0.3613)
(0.4366,0.3751)
(0.4380,0.3847)
(0.4393,0.3537)
(0.4407,0.3701)
(0.4420,0.3508)
(0.4433,0.3684)
(0.4446,0.3681)
(0.4459,0.3435)
(0.4471,0.3586)
(0.4483,0.3719)
(0.4495,0.3441)
(0.4507,0.3575)
(0.4518,0.3651)
(0.4530,0.3592)
(0.4541,0.3376)
(0.4552,0.3574)
(0.4563,0.3493)
(0.4573,0.3546)
(0.4584,0.3485)
(0.4594,0.3762)
(0.4604,0.3499)
(0.4614,0.3695)
(0.4624,0.3562)
(0.4634,0.3032)
(0.4644,0.3725)
(0.4653,0.3503)
(0.4663,0.3557)
(0.4672,0.3550)
(0.4681,0.3361)
(0.4690,0.3524)
(0.4699,0.3478)
(0.4708,0.3411)
(0.4716,0.3633)
(0.4725,0.3389)
(0.4733,0.3448)
(0.4742,0.3484)
(0.4750,0.3480)
(0.4758,0.3121)
(0.4766,0.3502)
(0.4774,0.3504)
(0.4782,0.3573)
(0.4790,0.3445)
(0.4798,0.3274)
(0.4805,0.3072)
(0.4813,0.3535)
(0.4820,0.3175)
(0.4828,0.3321)
(0.4835,0.3618)
(0.4842,0.3517)
(0.4849,0.3225)
(0.4856,0.3408)
(0.4863,0.3322)
\PST@Solid(0.4863,0.3322)
(0.4870,0.3488)
(0.4877,0.3403)
(0.4884,0.3408)
(0.4891,0.3392)
(0.4897,0.3398)
(0.4904,0.3555)
(0.4911,0.3363)
(0.4917,0.3480)
(0.4923,0.3441)
(0.4930,0.3468)
(0.4936,0.3484)
(0.4942,0.3017)
(0.4949,0.3343)
(0.4955,0.3148)
(0.4961,0.3428)
(0.4967,0.3539)
(0.4973,0.3410)
(0.4979,0.3278)
(0.4985,0.3427)
(0.4991,0.3410)
(0.4996,0.3279)
(0.5002,0.2512)
(0.5008,0.3368)
(0.5013,0.3427)
(0.5019,0.2841)
(0.5025,0.3301)
(0.5030,0.3404)
(0.5036,0.3454)
(0.5041,0.3047)
(0.5046,0.3273)
(0.5052,0.3572)
(0.5057,0.3299)
(0.5062,0.3264)
(0.5068,0.3421)
(0.5073,0.3260)
(0.5078,0.3454)
(0.5083,0.2820)
(0.5088,0.3284)
(0.5093,0.3309)
(0.5098,0.3426)
(0.5103,0.3436)
(0.5108,0.3452)
(0.5113,0.3505)
(0.5118,0.3192)
(0.5123,0.3028)
(0.5127,0.3424)
(0.5132,0.3368)
(0.5137,0.3293)
(0.5142,0.3281)
(0.5146,0.3363)
(0.5151,0.3404)
(0.5155,0.2689)
(0.5160,0.3145)
(0.5165,0.3348)
(0.5169,0.2976)
(0.5174,0.3192)
(0.5178,0.2808)
(0.5183,0.3366)
(0.5187,0.3184)
(0.5191,0.3264)
(0.5196,0.3249)
(0.5200,0.3111)
(0.5204,0.3452)
(0.5209,0.3134)
(0.5213,0.3266)
(0.5217,0.2906)
(0.5221,0.3155)
(0.5225,0.3453)
(0.5229,0.3372)
(0.5234,0.3240)
(0.5238,0.2969)
(0.5242,0.3086)
(0.5246,0.3222)
(0.5250,0.3262)
(0.5254,0.3148)
(0.5258,0.3169)
(0.5262,0.3325)
(0.5266,0.3317)
(0.5270,0.3182)
(0.5273,0.3359)
(0.5277,0.3089)
(0.5281,0.3365)
(0.5285,0.2570)
(0.5289,0.3229)
(0.5293,0.3384)
(0.5296,0.2982)
(0.5300,0.3235)
(0.5304,0.3288)
(0.5308,0.3345)
(0.5311,0.3128)
(0.5315,0.3150)
(0.5319,0.3167)
(0.5322,0.3286)
(0.5326,0.3322)
(0.5329,0.2900)
(0.5333,0.3200)
(0.5336,0.3405)
(0.5340,0.3334)
(0.5343,0.3350)
(0.5347,0.3119)
\PST@Solid(0.5347,0.3119)
(0.5350,0.2895)
(0.5354,0.3150)
(0.5357,0.2931)
(0.5361,0.3308)
(0.5364,0.3128)
(0.5368,0.2791)
(0.5371,0.3266)
(0.5374,0.3235)
(0.5378,0.3110)
(0.5381,0.3245)
(0.5384,0.3269)
(0.5388,0.3135)
(0.5391,0.2913)
(0.5394,0.2926)
(0.5397,0.3352)
(0.5401,0.3263)
(0.5404,0.2924)
(0.5407,0.2873)
(0.5410,0.3452)
(0.5413,0.3355)
(0.5417,0.3281)
(0.5420,0.3002)
(0.5423,0.3360)
(0.5426,0.3342)
(0.5429,0.2552)
(0.5432,0.3182)
(0.5435,0.3144)
(0.5438,0.3420)
(0.5441,0.3122)
(0.5445,0.3262)
(0.5448,0.3241)
(0.5451,0.3180)
(0.5454,0.3388)
(0.5457,0.2938)
(0.5460,0.2912)
(0.5462,0.3243)
(0.5465,0.2880)
(0.5468,0.3078)
(0.5471,0.3348)
(0.5474,0.3154)
(0.5477,0.3186)
(0.5480,0.3052)
(0.5483,0.3215)
(0.5486,0.2938)
(0.5489,0.3053)
(0.5491,0.3246)
(0.5494,0.2902)
(0.5497,0.3032)
(0.5500,0.2727)
(0.5503,0.3087)
(0.5505,0.3281)
(0.5508,0.3054)
(0.5511,0.3149)
(0.5514,0.3029)
(0.5517,0.3118)
(0.5519,0.2874)
(0.5522,0.3097)
(0.5525,0.3275)
(0.5527,0.3214)
(0.5530,0.3391)
(0.5533,0.2958)
(0.5535,0.2964)
(0.5538,0.3273)
(0.5541,0.2902)
(0.5543,0.3284)
(0.5546,0.3097)
(0.5549,0.2825)
(0.5551,0.2717)
(0.5554,0.3108)
(0.5556,0.2838)
(0.5559,0.3183)
(0.5562,0.2340)
(0.5564,0.2914)
(0.5567,0.3219)
(0.5569,0.2722)
(0.5572,0.2908)
(0.5574,0.3032)
(0.5577,0.3232)
(0.5579,0.3091)
(0.5582,0.3181)
(0.5584,0.3265)
(0.5587,0.3180)
(0.5589,0.2735)
(0.5592,0.3005)
(0.5594,0.2765)
(0.5597,0.3190)
(0.5599,0.3201)
(0.5601,0.2511)
(0.5604,0.2592)
(0.5606,0.2538)
(0.5609,0.3105)
(0.5611,0.3274)
(0.5613,0.2775)
(0.5616,0.3279)
(0.5618,0.2863)
(0.5621,0.2640)
(0.5623,0.2992)
(0.5625,0.2875)
(0.5628,0.2622)
(0.5630,0.3274)
\PST@Solid(0.5630,0.3274)
(0.5632,0.3239)
(0.5635,0.2963)
(0.5637,0.3041)
(0.5639,0.2934)
(0.5641,0.3083)
(0.5644,0.3167)
(0.5646,0.3351)
(0.5648,0.2789)
(0.5651,0.2485)
(0.5653,0.3100)
(0.5655,0.3117)
(0.5657,0.3108)
(0.5660,0.2858)
(0.5662,0.2922)
(0.5664,0.3033)
(0.5666,0.3050)
(0.5668,0.3127)
(0.5671,0.2127)
(0.5673,0.3092)
(0.5675,0.2999)
(0.5677,0.2997)
(0.5679,0.2964)
(0.5681,0.2820)
(0.5684,0.3271)
(0.5686,0.3203)
(0.5688,0.3143)
(0.5690,0.3411)
(0.5692,0.2817)
(0.5694,0.3131)
(0.5696,0.3155)
(0.5699,0.3234)
(0.5701,0.2795)
(0.5703,0.2984)
(0.5705,0.2865)
(0.5707,0.2877)
(0.5709,0.2342)
(0.5711,0.3141)
(0.5713,0.3241)
(0.5715,0.2937)
(0.5717,0.3323)
(0.5719,0.2839)
(0.5721,0.3174)
(0.5723,0.2672)
(0.5725,0.3181)
(0.5727,0.3215)
(0.5729,0.3268)
(0.5731,0.3275)
(0.5733,0.3135)
(0.5735,0.2977)
(0.5737,0.3059)
(0.5739,0.3395)
(0.5741,0.2844)
(0.5743,0.2370)
(0.5745,0.2844)
(0.5747,0.3395)
(0.5749,0.3059)
(0.5751,0.2977)
(0.5753,0.3135)
(0.5755,0.3275)
(0.5757,0.3268)
(0.5759,0.3215)
(0.5761,0.3181)
(0.5763,0.2672)
(0.5765,0.3174)
(0.5767,0.2839)
(0.5769,0.3323)
(0.5771,0.2937)
(0.5772,0.3241)
(0.5774,0.3141)
(0.5776,0.2342)
(0.5778,0.2877)
(0.5780,0.2865)
(0.5782,0.2984)
(0.5784,0.2795)
(0.5786,0.3234)
(0.5787,0.3155)
(0.5789,0.3131)
(0.5791,0.2817)
(0.5793,0.3411)
(0.5795,0.3143)
(0.5797,0.3203)
(0.5799,0.3271)
(0.5800,0.2820)
(0.5802,0.2964)
(0.5804,0.2997)
(0.5806,0.2999)
(0.5808,0.3092)
(0.5809,0.2127)
(0.5811,0.3127)
(0.5813,0.3050)
(0.5815,0.3033)
(0.5817,0.2922)
(0.5818,0.2858)
(0.5820,0.3108)
(0.5822,0.3117)
(0.5824,0.3100)
(0.5825,0.2485)
(0.5827,0.2789)
(0.5829,0.3351)
(0.5831,0.3167)
\PST@Solid(0.5831,0.3167)
(0.5832,0.3083)
(0.5834,0.2934)
(0.5836,0.3041)
(0.5838,0.2963)
(0.5839,0.3239)
(0.5841,0.3274)
(0.5843,0.2622)
(0.5844,0.2875)
(0.5846,0.2992)
(0.5848,0.2640)
(0.5850,0.2863)
(0.5851,0.3279)
(0.5853,0.2775)
(0.5855,0.3274)
(0.5856,0.3105)
(0.5858,0.2538)
(0.5860,0.2592)
(0.5861,0.2511)
(0.5863,0.3201)
(0.5865,0.3190)
(0.5866,0.2765)
(0.5868,0.3005)
(0.5870,0.2735)
(0.5871,0.3180)
(0.5873,0.3265)
(0.5875,0.3181)
(0.5876,0.3091)
(0.5878,0.3232)
(0.5879,0.3032)
(0.5881,0.2908)
(0.5883,0.2722)
(0.5884,0.3219)
(0.5886,0.2914)
(0.5888,0.2340)
(0.5889,0.3183)
(0.5891,0.2838)
(0.5892,0.3108)
(0.5894,0.2717)
(0.5896,0.2825)
(0.5897,0.3097)
(0.5899,0.3284)
(0.5900,0.2902)
(0.5902,0.3273)
(0.5903,0.2964)
(0.5905,0.2958)
(0.5907,0.3391)
(0.5908,0.3214)
(0.5910,0.3275)
(0.5911,0.3097)
(0.5913,0.2874)
(0.5914,0.3118)
(0.5916,0.3029)
(0.5917,0.3149)
(0.5919,0.3054)
(0.5921,0.3281)
(0.5922,0.3087)
(0.5924,0.2727)
(0.5925,0.3032)
(0.5927,0.2902)
(0.5928,0.3246)
(0.5930,0.3053)
(0.5931,0.2938)
(0.5933,0.3215)
(0.5934,0.3052)
(0.5936,0.3186)
(0.5937,0.3154)
(0.5939,0.3348)
(0.5940,0.3078)
(0.5942,0.2880)
(0.5943,0.3243)
(0.5945,0.2912)
(0.5946,0.2938)
(0.5948,0.3388)
(0.5949,0.3180)
(0.5951,0.3241)
(0.5952,0.3262)
(0.5953,0.3122)
(0.5955,0.3420)
(0.5956,0.3144)
(0.5958,0.3182)
(0.5959,0.2552)
(0.5961,0.3342)
(0.5962,0.3360)
(0.5964,0.3002)
(0.5965,0.3281)
(0.5967,0.3355)
(0.5968,0.3452)
(0.5969,0.2873)
(0.5971,0.2924)
(0.5972,0.3263)
(0.5974,0.3352)
(0.5975,0.2926)
(0.5977,0.2913)
(0.5978,0.3135)
(0.5979,0.3269)
(0.5981,0.3245)
(0.5982,0.3110)
(0.5984,0.3235)
(0.5985,0.3266)
(0.5986,0.2791)
\PST@Solid(0.5986,0.2791)
(0.5988,0.3128)
(0.5989,0.3308)
(0.5991,0.2931)
(0.5992,0.3150)
(0.5993,0.2895)
(0.5995,0.3119)
(0.5996,0.3350)
(0.5997,0.3334)
(0.5999,0.3405)
(0.6000,0.3200)
(0.6002,0.2900)
(0.6003,0.3322)
(0.6004,0.3286)
(0.6006,0.3167)
(0.6007,0.3150)
(0.6008,0.3128)
(0.6010,0.3345)
(0.6011,0.3288)
(0.6012,0.3235)
(0.6014,0.2982)
(0.6015,0.3384)
(0.6016,0.3229)
(0.6018,0.2570)
(0.6019,0.3365)
(0.6020,0.3089)
(0.6022,0.3359)
(0.6023,0.3182)
(0.6024,0.3317)
(0.6026,0.3325)
(0.6027,0.3169)
(0.6028,0.3148)
(0.6030,0.3262)
(0.6031,0.3222)
(0.6032,0.3086)
(0.6034,0.2969)
(0.6035,0.3240)
(0.6036,0.3372)
(0.6037,0.3453)
(0.6039,0.3155)
(0.6040,0.2906)
(0.6041,0.3266)
(0.6043,0.3134)
(0.6044,0.3452)
(0.6045,0.3111)
(0.6046,0.3249)
(0.6048,0.3264)
(0.6049,0.3184)
(0.6050,0.3366)
(0.6052,0.2808)
(0.6053,0.3192)
(0.6054,0.2976)
(0.6055,0.3348)
(0.6057,0.3145)
(0.6058,0.2689)
(0.6059,0.3404)
(0.6060,0.3363)
(0.6062,0.3281)
(0.6063,0.3293)
(0.6064,0.3368)
(0.6065,0.3424)
(0.6067,0.3028)
(0.6068,0.3192)
(0.6069,0.3505)
(0.6070,0.3452)
(0.6072,0.3436)
(0.6073,0.3426)
(0.6074,0.3309)
(0.6075,0.3284)
(0.6077,0.2820)
(0.6078,0.3454)
(0.6079,0.3260)
(0.6080,0.3421)
(0.6081,0.3264)
(0.6083,0.3299)
(0.6084,0.3572)
(0.6085,0.3273)
(0.6086,0.3047)
(0.6087,0.3454)
(0.6089,0.3404)
(0.6090,0.3301)
(0.6091,0.2841)
(0.6092,0.3427)
(0.6094,0.3368)
(0.6095,0.2512)
(0.6096,0.3279)
(0.6097,0.3410)
(0.6098,0.3427)
(0.6099,0.3278)
(0.6101,0.3410)
(0.6102,0.3539)
(0.6103,0.3428)
(0.6104,0.3148)
(0.6105,0.3343)
(0.6107,0.3017)
(0.6108,0.3484)
(0.6109,0.3468)
(0.6110,0.3441)
(0.6111,0.3480)
(0.6112,0.3363)
(0.6114,0.3555)
\PST@Solid(0.6114,0.3555)
(0.6115,0.3398)
(0.6116,0.3392)
(0.6117,0.3408)
(0.6118,0.3403)
(0.6119,0.3488)
(0.6121,0.3322)
(0.6122,0.3408)
(0.6123,0.3225)
(0.6124,0.3517)
(0.6125,0.3618)
(0.6126,0.3321)
(0.6127,0.3175)
(0.6129,0.3535)
(0.6130,0.3072)
(0.6131,0.3274)
(0.6132,0.3445)
(0.6133,0.3573)
(0.6134,0.3504)
(0.6135,0.3502)
(0.6136,0.3121)
(0.6138,0.3480)
(0.6139,0.3484)
(0.6140,0.3448)
(0.6141,0.3389)
(0.6142,0.3633)
(0.6143,0.3411)
(0.6144,0.3478)
(0.6145,0.3524)
(0.6146,0.3361)
(0.6148,0.3550)
(0.6149,0.3557)
(0.6150,0.3503)
(0.6151,0.3725)
(0.6152,0.3032)
(0.6153,0.3562)
(0.6154,0.3695)
(0.6155,0.3499)
(0.6156,0.3762)
(0.6158,0.3485)
(0.6159,0.3546)
(0.6160,0.3493)
(0.6161,0.3574)
(0.6162,0.3376)
(0.6163,0.3592)
(0.6164,0.3651)
(0.6165,0.3575)
(0.6166,0.3441)
(0.6167,0.3719)
(0.6168,0.3586)
(0.6169,0.3435)
(0.6170,0.3681)
(0.6172,0.3684)
(0.6173,0.3508)
(0.6174,0.3701)
(0.6175,0.3537)
(0.6176,0.3847)
(0.6177,0.3751)
(0.6178,0.3613)
(0.6179,0.3637)
(0.6180,0.3465)
(0.6181,0.3704)
(0.6182,0.3594)
(0.6183,0.3648)
(0.6184,0.3912)
(0.6185,0.3663)
(0.6186,0.3579)
(0.6187,0.3909)
(0.6188,0.3860)
(0.6190,0.3448)
(0.6191,0.3716)
(0.6192,0.3910)
(0.6193,0.3915)
(0.6194,0.3156)
(0.6195,0.3693)
(0.6196,0.3946)
(0.6197,0.3704)
(0.6198,0.3787)
(0.6199,0.3827)
(0.6200,0.3902)
(0.6201,0.3898)
(0.6202,0.3640)
(0.6203,0.3790)
(0.6204,0.4070)
(0.6205,0.4023)
(0.6206,0.3723)
(0.6207,0.4098)
(0.6208,0.4132)
(0.6209,0.4005)
(0.6210,0.3899)
(0.6211,0.4133)
(0.6212,0.4211)
(0.6213,0.4014)
(0.6214,0.4082)
(0.6215,0.4424)
(0.6216,0.4288)
(0.6217,0.3928)
(0.6218,0.4046)
(0.6219,0.4354)
(0.6220,0.4383)
(0.6221,0.4377)
\PST@Solid(0.6221,0.4377)
(0.6222,0.4543)
(0.6223,0.4614)
(0.6224,0.4724)
(0.6225,0.4535)
(0.6226,0.5168)

\rput[r](0.5200,0.4850){$2.5\times10^4x^{-1}$}
\PST@Dashed(0.5360,0.4850)
(0.6150,0.4850)

\PST@Dashed(0.1650,0.5148)
(0.1650,0.5148)
(0.1696,0.5122)
(0.1742,0.5097)
(0.1789,0.5071)
(0.1835,0.5046)
(0.1881,0.5021)
(0.1927,0.4995)
(0.1974,0.4970)
(0.2020,0.4944)
(0.2066,0.4919)
(0.2112,0.4893)
(0.2158,0.4868)
(0.2205,0.4843)
(0.2251,0.4817)
(0.2297,0.4792)
(0.2343,0.4766)
(0.2390,0.4741)
(0.2436,0.4715)
(0.2482,0.4690)
(0.2528,0.4665)
(0.2574,0.4639)
(0.2621,0.4614)
(0.2667,0.4588)
(0.2713,0.4563)
(0.2759,0.4537)
(0.2806,0.4512)
(0.2852,0.4487)
(0.2898,0.4461)
(0.2944,0.4436)
(0.2990,0.4410)
(0.3037,0.4385)
(0.3083,0.4359)
(0.3129,0.4334)
(0.3175,0.4309)
(0.3222,0.4283)
(0.3268,0.4258)
(0.3314,0.4232)
(0.3360,0.4207)
(0.3406,0.4181)
(0.3453,0.4156)
(0.3499,0.4130)
(0.3545,0.4105)
(0.3591,0.4080)
(0.3638,0.4054)
(0.3684,0.4029)
(0.3730,0.4003)
(0.3776,0.3978)
(0.3822,0.3952)
(0.3869,0.3927)
(0.3915,0.3902)
(0.3961,0.3876)
(0.4007,0.3851)
(0.4054,0.3825)
(0.4100,0.3800)
(0.4146,0.3774)
(0.4192,0.3749)
(0.4238,0.3724)
(0.4285,0.3698)
(0.4331,0.3673)
(0.4377,0.3647)
(0.4423,0.3622)
(0.4470,0.3596)
(0.4516,0.3571)
(0.4562,0.3546)
(0.4608,0.3520)
(0.4655,0.3495)
(0.4701,0.3469)
(0.4747,0.3444)
(0.4793,0.3418)
(0.4839,0.3393)
(0.4886,0.3368)
(0.4932,0.3342)
(0.4978,0.3317)
(0.5024,0.3291)
(0.5071,0.3266)
(0.5117,0.3240)
(0.5163,0.3215)
(0.5209,0.3189)
(0.5255,0.3164)
(0.5302,0.3139)
(0.5348,0.3113)
(0.5394,0.3088)
(0.5440,0.3062)
(0.5487,0.3037)
(0.5533,0.3011)
(0.5579,0.2986)
(0.5625,0.2961)
(0.5671,0.2935)
(0.5718,0.2910)
(0.5764,0.2884)
(0.5810,0.2859)
(0.5856,0.2833)
(0.5903,0.2808)
(0.5949,0.2783)
(0.5995,0.2757)
(0.6041,0.2732)
(0.6087,0.2706)
(0.6134,0.2681)
(0.6180,0.2655)
(0.6226,0.2630)

\catcode`@=12
\fi
\endpspicture

%% file: webdivFFT.tex
\ifx\PSTloaded\undefined
\def\PSTloaded{t}
\psset{arrowsize=.01 3.2 1.4 .3}
\psset{dotsize=.01}
\catcode`@=11

\newpsobject{PST@Border}{psline}{linewidth=.0015,linestyle=solid}
\newpsobject{PST@Axes}{psline}{linewidth=.0015,linestyle=dotted,dotsep=.004}
\newpsobject{PST@Solid}{psline}{linewidth=.0015,linestyle=solid}
\newpsobject{PST@Dashed}{psline}{linewidth=.0015,linestyle=dashed,dash=.01 .01}
\newpsobject{PST@Dotted}{psline}{linewidth=.0025,linestyle=dotted,dotsep=.008}
\newpsobject{PST@LongDash}{psline}{linewidth=.0015,linestyle=dashed,dash=.02 .01}
\newpsobject{PST@Diamond}{psdots}{linewidth=.001,linestyle=solid,dotstyle=square,dotangle=45}
\newpsobject{PST@Filldiamond}{psdots}{linewidth=.001,linestyle=solid,dotstyle=square*,dotangle=45}
\newpsobject{PST@Cross}{psdots}{linewidth=.001,linestyle=solid,dotstyle=+,dotangle=45}
\newpsobject{PST@Plus}{psdots}{linewidth=.001,linestyle=solid,dotstyle=+}
\newpsobject{PST@Square}{psdots}{linewidth=.001,linestyle=solid,dotstyle=square}
\newpsobject{PST@Circle}{psdots}{linewidth=.001,linestyle=solid,dotstyle=o}
\newpsobject{PST@Triangle}{psdots}{linewidth=.001,linestyle=solid,dotstyle=triangle}
\newpsobject{PST@Pentagon}{psdots}{linewidth=.001,linestyle=solid,dotstyle=pentagon}
\newpsobject{PST@Fillsquare}{psdots}{linewidth=.001,linestyle=solid,dotstyle=square*}
\newpsobject{PST@Fillcircle}{psdots}{linewidth=.001,linestyle=solid,dotstyle=*}
\newpsobject{PST@Filltriangle}{psdots}{linewidth=.001,linestyle=solid,dotstyle=triangle*}
\newpsobject{PST@Fillpentagon}{psdots}{linewidth=.001,linestyle=solid,dotstyle=pentagon*}
\newpsobject{PST@Arrow}{psline}{linewidth=.001,linestyle=solid}
\catcode`@=12

\fi
\psset{unit=5.0in,xunit=5.0in,yunit=3.0in}
\pspicture(0.000000,0.000000)(0.700000,0.600000)
\ifx\nofigs\undefined
\catcode`@=11

\PST@Border(0.1490,0.1260)
(0.1640,0.1260)

\PST@Border(0.6470,0.1260)
(0.6320,0.1260)

\rput[r](0.1330,0.1260){ 0.1}
\PST@Border(0.1490,0.1526)
(0.1565,0.1526)

\PST@Border(0.6470,0.1526)
(0.6395,0.1526)

\PST@Border(0.1490,0.1878)
(0.1565,0.1878)

\PST@Border(0.6470,0.1878)
(0.6395,0.1878)

\PST@Border(0.1490,0.2058)
(0.1565,0.2058)

\PST@Border(0.6470,0.2058)
(0.6395,0.2058)

\PST@Border(0.1490,0.2144)
(0.1640,0.2144)

\PST@Border(0.6470,0.2144)
(0.6320,0.2144)

\rput[r](0.1330,0.2144){ 1}
\PST@Border(0.1490,0.2410)
(0.1565,0.2410)

\PST@Border(0.6470,0.2410)
(0.6395,0.2410)

\PST@Border(0.1490,0.2762)
(0.1565,0.2762)

\PST@Border(0.6470,0.2762)
(0.6395,0.2762)

\PST@Border(0.1490,0.2942)
(0.1565,0.2942)

\PST@Border(0.6470,0.2942)
(0.6395,0.2942)

\PST@Border(0.1490,0.3028)
(0.1640,0.3028)

\PST@Border(0.6470,0.3028)
(0.6320,0.3028)

\rput[r](0.1330,0.3028){ 10}
\PST@Border(0.1490,0.3294)
(0.1565,0.3294)

\PST@Border(0.6470,0.3294)
(0.6395,0.3294)

\PST@Border(0.1490,0.3646)
(0.1565,0.3646)

\PST@Border(0.6470,0.3646)
(0.6395,0.3646)

\PST@Border(0.1490,0.3826)
(0.1565,0.3826)

\PST@Border(0.6470,0.3826)
(0.6395,0.3826)

\PST@Border(0.1490,0.3912)
(0.1640,0.3912)

\PST@Border(0.6470,0.3912)
(0.6320,0.3912)

\rput[r](0.1330,0.3912){ 100}
\PST@Border(0.1490,0.4178)
(0.1565,0.4178)

\PST@Border(0.6470,0.4178)
(0.6395,0.4178)

\PST@Border(0.1490,0.4530)
(0.1565,0.4530)

\PST@Border(0.6470,0.4530)
(0.6395,0.4530)

\PST@Border(0.1490,0.4710)
(0.1565,0.4710)

\PST@Border(0.6470,0.4710)
(0.6395,0.4710)

\PST@Border(0.1490,0.4796)
(0.1640,0.4796)

\PST@Border(0.6470,0.4796)
(0.6320,0.4796)

\rput[r](0.1330,0.4796){ 1000}
\PST@Border(0.1490,0.5062)
(0.1565,0.5062)

\PST@Border(0.6470,0.5062)
(0.6395,0.5062)

\PST@Border(0.1490,0.5414)
(0.1565,0.5414)

\PST@Border(0.6470,0.5414)
(0.6395,0.5414)

\PST@Border(0.1490,0.5594)
(0.1565,0.5594)

\PST@Border(0.6470,0.5594)
(0.6395,0.5594)

\PST@Border(0.1490,0.5680)
(0.1640,0.5680)

\PST@Border(0.6470,0.5680)
(0.6320,0.5680)

\rput[r](0.1330,0.5680){ 10000}
\PST@Border(0.1490,0.1260)
(0.1490,0.1460)

\PST@Border(0.1490,0.5680)
(0.1490,0.5480)

\rput(0.1490,0.0840){ 1}
\PST@Border(0.1990,0.1260)
(0.1990,0.1360)

\PST@Border(0.1990,0.5680)
(0.1990,0.5580)

\PST@Border(0.2282,0.1260)
(0.2282,0.1360)

\PST@Border(0.2282,0.5680)
(0.2282,0.5580)

\PST@Border(0.2489,0.1260)
(0.2489,0.1360)

\PST@Border(0.2489,0.5680)
(0.2489,0.5580)

\PST@Border(0.2650,0.1260)
(0.2650,0.1360)

\PST@Border(0.2650,0.5680)
(0.2650,0.5580)

\PST@Border(0.2782,0.1260)
(0.2782,0.1360)

\PST@Border(0.2782,0.5680)
(0.2782,0.5580)

\PST@Border(0.2893,0.1260)
(0.2893,0.1360)

\PST@Border(0.2893,0.5680)
(0.2893,0.5580)

\PST@Border(0.2989,0.1260)
(0.2989,0.1360)

\PST@Border(0.2989,0.5680)
(0.2989,0.5580)

\PST@Border(0.3074,0.1260)
(0.3074,0.1360)

\PST@Border(0.3074,0.5680)
(0.3074,0.5580)

\PST@Border(0.3150,0.1260)
(0.3150,0.1460)

\PST@Border(0.3150,0.5680)
(0.3150,0.5480)

\rput(0.3150,0.0840){ 10}
\PST@Border(0.3650,0.1260)
(0.3650,0.1360)

\PST@Border(0.3650,0.5680)
(0.3650,0.5580)

\PST@Border(0.3942,0.1260)
(0.3942,0.1360)

\PST@Border(0.3942,0.5680)
(0.3942,0.5580)

\PST@Border(0.4149,0.1260)
(0.4149,0.1360)

\PST@Border(0.4149,0.5680)
(0.4149,0.5580)

\PST@Border(0.4310,0.1260)
(0.4310,0.1360)

\PST@Border(0.4310,0.5680)
(0.4310,0.5580)

\PST@Border(0.4442,0.1260)
(0.4442,0.1360)

\PST@Border(0.4442,0.5680)
(0.4442,0.5580)

\PST@Border(0.4553,0.1260)
(0.4553,0.1360)

\PST@Border(0.4553,0.5680)
(0.4553,0.5580)

\PST@Border(0.4649,0.1260)
(0.4649,0.1360)

\PST@Border(0.4649,0.5680)
(0.4649,0.5580)

\PST@Border(0.4734,0.1260)
(0.4734,0.1360)

\PST@Border(0.4734,0.5680)
(0.4734,0.5580)

\PST@Border(0.4810,0.1260)
(0.4810,0.1460)

\PST@Border(0.4810,0.5680)
(0.4810,0.5480)

\rput(0.4810,0.0840){ 100}
\PST@Border(0.5310,0.1260)
(0.5310,0.1360)

\PST@Border(0.5310,0.5680)
(0.5310,0.5580)

\PST@Border(0.5602,0.1260)
(0.5602,0.1360)

\PST@Border(0.5602,0.5680)
(0.5602,0.5580)

\PST@Border(0.5809,0.1260)
(0.5809,0.1360)

\PST@Border(0.5809,0.5680)
(0.5809,0.5580)

\PST@Border(0.5970,0.1260)
(0.5970,0.1360)

\PST@Border(0.5970,0.5680)
(0.5970,0.5580)

\PST@Border(0.6102,0.1260)
(0.6102,0.1360)

\PST@Border(0.6102,0.5680)
(0.6102,0.5580)

\PST@Border(0.6213,0.1260)
(0.6213,0.1360)

\PST@Border(0.6213,0.5680)
(0.6213,0.5580)

\PST@Border(0.6309,0.1260)
(0.6309,0.1360)

\PST@Border(0.6309,0.5680)
(0.6309,0.5580)

\PST@Border(0.6394,0.1260)
(0.6394,0.1360)

\PST@Border(0.6394,0.5680)
(0.6394,0.5580)

\PST@Border(0.6470,0.1260)
(0.6470,0.1460)

\PST@Border(0.6470,0.5680)
(0.6470,0.5480)

\rput(0.6470,0.0840){ 1000}
\PST@Border(0.1490,0.1260)
(0.6470,0.1260)
(0.6470,0.5680)
(0.1490,0.5680)
(0.1490,0.1260)

\rput(0.3980,0.0210){$k$}
\rput[r](0.5200,0.5270){$|\tilde{D}(k)|$}
\PST@Solid(0.5360,0.5270)
(0.6150,0.5270)

\PST@Solid(0.1490,0.5253)
(0.1490,0.5253)
(0.1990,0.5132)
(0.2282,0.4541)
(0.2489,0.4807)
(0.2650,0.4774)
(0.2782,0.4415)
(0.2893,0.4538)
(0.2989,0.4476)
(0.3074,0.4432)
(0.3150,0.4697)
(0.3219,0.3966)
(0.3281,0.4448)
(0.3339,0.4249)
(0.3393,0.4295)
(0.3442,0.4355)
(0.3489,0.4459)
(0.3533,0.3162)
(0.3574,0.4323)
(0.3613,0.4417)
(0.3650,0.4106)
(0.3685,0.4046)
(0.3718,0.4063)
(0.3750,0.4059)
(0.3781,0.3929)
(0.3811,0.4254)
(0.3839,0.4134)
(0.3866,0.3893)
(0.3892,0.3231)
(0.3918,0.3775)
(0.3942,0.4074)
(0.3966,0.3973)
(0.3989,0.4092)
(0.4011,0.3810)
(0.4032,0.3727)
(0.4053,0.3771)
(0.4073,0.4062)
(0.4093,0.4029)
(0.4112,0.4175)
(0.4131,0.4061)
(0.4149,0.3461)
(0.4167,0.3887)
(0.4185,0.3848)
(0.4202,0.4096)
(0.4218,0.4192)
(0.4234,0.3654)
(0.4250,0.3798)
(0.4266,0.3659)
(0.4281,0.3673)
(0.4296,0.4008)
(0.4310,0.3702)
(0.4325,0.3646)
(0.4339,0.3559)
(0.4352,0.3635)
(0.4366,0.4026)
(0.4379,0.3512)
(0.4392,0.3547)
(0.4405,0.3591)
(0.4417,0.3804)
(0.4430,0.3898)
(0.4442,0.3819)
(0.4454,0.3722)
(0.4465,0.3914)
(0.4477,0.3295)
(0.4488,0.3465)
(0.4499,0.3193)
(0.4510,0.3544)
(0.4521,0.3257)
(0.4532,0.3724)
(0.4542,0.3716)
(0.4553,0.3494)
(0.4563,0.3843)
(0.4573,0.3660)
(0.4583,0.3612)
(0.4593,0.3748)
(0.4603,0.2848)
(0.4612,0.3587)
(0.4622,0.3712)
(0.4631,0.3578)
(0.4640,0.3529)
(0.4649,0.3428)
(0.4658,0.2998)
(0.4667,0.3767)
(0.4676,0.3510)
(0.4684,0.3680)
(0.4693,0.3675)
(0.4701,0.3807)
(0.4710,0.3701)
(0.4718,0.3482)
(0.4726,0.3380)
(0.4734,0.3844)
(0.4742,0.3543)
(0.4750,0.3781)
(0.4758,0.3587)
(0.4765,0.2832)
(0.4773,0.3517)
(0.4781,0.3616)
(0.4788,0.3089)
(0.4795,0.3066)
(0.4803,0.3222)
(0.4810,0.3574)
\PST@Solid(0.4810,0.3574)
(0.4817,0.3461)
(0.4824,0.3429)
(0.4831,0.3423)
(0.4838,0.3571)
(0.4845,0.3534)
(0.4852,0.3100)
(0.4859,0.3521)
(0.4865,0.3542)
(0.4872,0.2810)
(0.4879,0.3502)
(0.4885,0.3600)
(0.4892,0.3357)
(0.4898,0.3523)
(0.4904,0.3403)
(0.4911,0.3445)
(0.4917,0.3707)
(0.4923,0.3434)
(0.4929,0.3195)
(0.4935,0.3335)
(0.4941,0.3574)
(0.4947,0.3413)
(0.4953,0.3553)
(0.4959,0.2894)
(0.4965,0.3502)
(0.4971,0.3611)
(0.4977,0.3434)
(0.4982,0.3524)
(0.4988,0.3422)
(0.4994,0.3142)
(0.4999,0.3602)
(0.5005,0.3422)
(0.5010,0.3408)
(0.5016,0.3152)
(0.5021,0.3532)
(0.5026,0.3518)
(0.5032,0.3469)
(0.5037,0.3464)
(0.5042,0.3585)
(0.5047,0.3155)
(0.5053,0.2967)
(0.5058,0.3600)
(0.5063,0.3556)
(0.5068,0.3081)
(0.5073,0.3096)
(0.5078,0.3667)
(0.5083,0.3637)
(0.5088,0.3529)
(0.5093,0.3389)
(0.5097,0.2984)
(0.5102,0.2681)
(0.5107,0.2850)
(0.5112,0.3335)
(0.5117,0.3432)
(0.5121,0.3090)
(0.5126,0.3146)
(0.5131,0.3123)
(0.5135,0.3216)
(0.5140,0.2715)
(0.5144,0.3552)
(0.5149,0.3268)
(0.5153,0.3420)
(0.5158,0.3441)
(0.5162,0.3538)
(0.5167,0.3449)
(0.5171,0.2981)
(0.5175,0.2582)
(0.5180,0.3251)
(0.5184,0.2997)
(0.5188,0.3350)
(0.5193,0.3516)
(0.5197,0.3431)
(0.5201,0.3272)
(0.5205,0.3228)
(0.5209,0.2958)
(0.5213,0.3353)
(0.5218,0.3143)
(0.5222,0.3042)
(0.5226,0.2732)
(0.5230,0.3255)
(0.5234,0.3453)
(0.5238,0.3336)
(0.5242,0.3018)
(0.5246,0.3414)
(0.5250,0.3402)
(0.5254,0.2373)
(0.5257,0.3096)
(0.5261,0.3508)
(0.5265,0.3048)
(0.5269,0.3333)
(0.5273,0.3197)
(0.5277,0.3571)
(0.5280,0.3082)
(0.5284,0.3431)
(0.5288,0.3272)
(0.5291,0.2929)
(0.5295,0.3346)
(0.5299,0.2871)
(0.5302,0.3073)
(0.5306,0.3604)
(0.5310,0.3489)
\PST@Solid(0.5310,0.3489)
(0.5313,0.3194)
(0.5317,0.3043)
(0.5320,0.3090)
(0.5324,0.3092)
(0.5328,0.3074)
(0.5331,0.3394)
(0.5335,0.3237)
(0.5338,0.3065)
(0.5341,0.3381)
(0.5345,0.3040)
(0.5348,0.3450)
(0.5352,0.3401)
(0.5355,0.3196)
(0.5358,0.3363)
(0.5362,0.3284)
(0.5365,0.2989)
(0.5369,0.3171)
(0.5372,0.3150)
(0.5375,0.2785)
(0.5378,0.2907)
(0.5382,0.3373)
(0.5385,0.3504)
(0.5388,0.3067)
(0.5391,0.3418)
(0.5395,0.3108)
(0.5398,0.3519)
(0.5401,0.2553)
(0.5404,0.3220)
(0.5407,0.3136)
(0.5410,0.2866)
(0.5414,0.3456)
(0.5417,0.3373)
(0.5420,0.2995)
(0.5423,0.3196)
(0.5426,0.3431)
(0.5429,0.3471)
(0.5432,0.2640)
(0.5435,0.3333)
(0.5438,0.2949)
(0.5441,0.3125)
(0.5444,0.3490)
(0.5447,0.2891)
(0.5450,0.3247)
(0.5453,0.3443)
(0.5456,0.3316)
(0.5459,0.3159)
(0.5462,0.3047)
(0.5465,0.3435)
(0.5468,0.3286)
(0.5471,0.3254)
(0.5473,0.3115)
(0.5476,0.3064)
(0.5479,0.3327)
(0.5482,0.2598)
(0.5485,0.2689)
(0.5488,0.3302)
(0.5490,0.3198)
(0.5493,0.3109)
(0.5496,0.3080)
(0.5499,0.2898)
(0.5502,0.2858)
(0.5504,0.3238)
(0.5507,0.3330)
(0.5510,0.3293)
(0.5513,0.3314)
(0.5515,0.3219)
(0.5518,0.3123)
(0.5521,0.2979)
(0.5523,0.3412)
(0.5526,0.2948)
(0.5529,0.3044)
(0.5531,0.3282)
(0.5534,0.3140)
(0.5537,0.3397)
(0.5539,0.3289)
(0.5542,0.3183)
(0.5545,0.3157)
(0.5547,0.3211)
(0.5550,0.3151)
(0.5552,0.2751)
(0.5555,0.3232)
(0.5557,0.3082)
(0.5560,0.3276)
(0.5563,0.3184)
(0.5565,0.3178)
(0.5568,0.2094)
(0.5570,0.2263)
(0.5573,0.2739)
(0.5575,0.3186)
(0.5578,0.3217)
(0.5580,0.2943)
(0.5583,0.2506)
(0.5585,0.2843)
(0.5587,0.2936)
(0.5590,0.3329)
(0.5592,0.3181)
(0.5595,0.3092)
(0.5597,0.3356)
(0.5600,0.3197)
(0.5602,0.3126)
\PST@Solid(0.5602,0.3126)
(0.5604,0.3296)
(0.5607,0.3291)
(0.5609,0.3213)
(0.5612,0.3280)
(0.5614,0.2627)
(0.5616,0.3223)
(0.5619,0.3191)
(0.5621,0.3073)
(0.5623,0.2670)
(0.5626,0.3353)
(0.5628,0.3226)
(0.5630,0.3412)
(0.5633,0.2888)
(0.5635,0.2697)
(0.5637,0.3183)
(0.5639,0.3155)
(0.5642,0.2718)
(0.5644,0.2783)
(0.5646,0.3360)
(0.5649,0.3116)
(0.5651,0.3143)
(0.5653,0.2951)
(0.5655,0.2896)
(0.5658,0.2850)
(0.5660,0.3300)
(0.5662,0.3113)
(0.5664,0.2824)
(0.5666,0.2861)
(0.5669,0.3144)
(0.5671,0.3272)
(0.5673,0.3221)
(0.5675,0.2575)
(0.5677,0.3237)
(0.5679,0.2671)
(0.5682,0.2907)
(0.5684,0.3120)
(0.5686,0.2997)
(0.5688,0.2281)
(0.5690,0.2865)
(0.5692,0.3173)
(0.5694,0.3107)
(0.5696,0.3233)
(0.5699,0.3089)
(0.5701,0.3156)
(0.5703,0.2904)
(0.5705,0.3301)
(0.5707,0.3201)
(0.5709,0.2977)
(0.5711,0.2621)
(0.5713,0.2781)
(0.5715,0.2944)
(0.5717,0.2952)
(0.5719,0.3334)
(0.5721,0.3061)
(0.5723,0.3095)
(0.5725,0.3292)
(0.5727,0.2827)
(0.5729,0.3475)
(0.5731,0.3159)
(0.5733,0.3212)
(0.5735,0.2995)
(0.5737,0.2907)
(0.5739,0.3109)
(0.5741,0.3204)
(0.5743,0.2684)
(0.5745,0.3152)
(0.5747,0.2908)
(0.5749,0.3200)
(0.5751,0.2860)
(0.5753,0.2970)
(0.5755,0.2698)
(0.5757,0.2913)
(0.5759,0.2972)
(0.5761,0.3211)
(0.5763,0.3282)
(0.5765,0.3191)
(0.5767,0.3057)
(0.5769,0.3346)
(0.5771,0.3042)
(0.5772,0.2730)
(0.5774,0.3219)
(0.5776,0.2544)
(0.5778,0.2755)
(0.5780,0.3265)
(0.5782,0.2643)
(0.5784,0.2378)
(0.5786,0.2833)
(0.5787,0.3209)
(0.5789,0.2931)
(0.5791,0.3196)
(0.5793,0.3203)
(0.5795,0.2567)
(0.5797,0.3333)
(0.5799,0.2980)
(0.5800,0.3327)
(0.5802,0.3088)
(0.5804,0.3086)
(0.5806,0.3119)
(0.5808,0.2885)
(0.5809,0.2815)
\PST@Solid(0.5809,0.2815)
(0.5811,0.3242)
(0.5813,0.3345)
(0.5815,0.2984)
(0.5817,0.2882)
(0.5818,0.3139)
(0.5820,0.3177)
(0.5822,0.3032)
(0.5824,0.3236)
(0.5825,0.3222)
(0.5827,0.3176)
(0.5829,0.2900)
(0.5831,0.2776)
(0.5832,0.2887)
(0.5834,0.3255)
(0.5836,0.2699)
(0.5838,0.2822)
(0.5839,0.2604)
(0.5841,0.2650)
(0.5843,0.3013)
(0.5845,0.3380)
(0.5846,0.3369)
(0.5848,0.3190)
(0.5850,0.3218)
(0.5851,0.2805)
(0.5853,0.3023)
(0.5855,0.3263)
(0.5857,0.3216)
(0.5858,0.3296)
(0.5860,0.3142)
(0.5862,0.3213)
(0.5863,0.2729)
(0.5865,0.2907)
(0.5867,0.3162)
(0.5868,0.3119)
(0.5870,0.3292)
(0.5872,0.3197)
(0.5873,0.3030)
(0.5875,0.3277)
(0.5876,0.2980)
(0.5878,0.3155)
(0.5880,0.2857)
(0.5881,0.2871)
(0.5883,0.2732)
(0.5885,0.2835)
(0.5886,0.3188)
(0.5888,0.2925)
(0.5890,0.1784)
(0.5891,0.3004)
(0.5893,0.2951)
(0.5894,0.2951)
(0.5896,0.3023)
(0.5898,0.3037)
(0.5899,0.2908)
(0.5901,0.3039)
(0.5902,0.3114)
(0.5904,0.3097)
(0.5905,0.3211)
(0.5907,0.3273)
(0.5909,0.3223)
(0.5910,0.3170)
(0.5912,0.2680)
(0.5913,0.3232)
(0.5915,0.2922)
(0.5916,0.3264)
(0.5918,0.2989)
(0.5920,0.2918)
(0.5921,0.3212)
(0.5923,0.3067)
(0.5924,0.3179)
(0.5926,0.3006)
(0.5927,0.3104)
(0.5929,0.3112)
(0.5930,0.3345)
(0.5932,0.2925)
(0.5933,0.3159)
(0.5935,0.2593)
(0.5936,0.2957)
(0.5938,0.3286)
(0.5939,0.3196)
(0.5941,0.2983)
(0.5942,0.2840)
(0.5944,0.3170)
(0.5945,0.2815)
(0.5947,0.3220)
(0.5948,0.3368)
(0.5950,0.2966)
(0.5951,0.2506)
(0.5953,0.2799)
(0.5954,0.3058)
(0.5956,0.3198)
(0.5957,0.3010)
(0.5959,0.3374)
(0.5960,0.3221)
(0.5962,0.3038)
(0.5963,0.3195)
(0.5964,0.2991)
(0.5966,0.3316)
(0.5967,0.2940)
(0.5969,0.2913)
(0.5970,0.2988)
\PST@Solid(0.5970,0.2988)
(0.5972,0.2913)
(0.5973,0.2940)
(0.5975,0.3316)
(0.5976,0.2991)
(0.5977,0.3195)
(0.5979,0.3038)
(0.5980,0.3221)
(0.5982,0.3374)
(0.5983,0.3010)
(0.5985,0.3198)
(0.5986,0.3058)
(0.5987,0.2799)
(0.5989,0.2506)
(0.5990,0.2966)
(0.5992,0.3368)
(0.5993,0.3220)
(0.5994,0.2815)
(0.5996,0.3170)
(0.5997,0.2840)
(0.5999,0.2983)
(0.6000,0.3196)
(0.6001,0.3286)
(0.6003,0.2957)
(0.6004,0.2593)
(0.6005,0.3159)
(0.6007,0.2925)
(0.6008,0.3345)
(0.6010,0.3112)
(0.6011,0.3104)
(0.6012,0.3006)
(0.6014,0.3179)
(0.6015,0.3067)
(0.6016,0.3212)
(0.6018,0.2918)
(0.6019,0.2989)
(0.6020,0.3264)
(0.6022,0.2922)
(0.6023,0.3232)
(0.6024,0.2680)
(0.6026,0.3170)
(0.6027,0.3223)
(0.6028,0.3273)
(0.6030,0.3211)
(0.6031,0.3097)
(0.6032,0.3114)
(0.6034,0.3039)
(0.6035,0.2908)
(0.6036,0.3037)
(0.6038,0.3023)
(0.6039,0.2951)
(0.6040,0.2951)
(0.6042,0.3004)
(0.6043,0.1784)
(0.6044,0.2925)
(0.6046,0.3188)
(0.6047,0.2835)
(0.6048,0.2732)
(0.6049,0.2871)
(0.6051,0.2857)
(0.6052,0.3155)
(0.6053,0.2980)
(0.6055,0.3277)
(0.6056,0.3030)
(0.6057,0.3197)
(0.6058,0.3292)
(0.6060,0.3119)
(0.6061,0.3162)
(0.6062,0.2907)
(0.6063,0.2729)
(0.6065,0.3213)
(0.6066,0.3142)
(0.6067,0.3296)
(0.6069,0.3216)
(0.6070,0.3263)
(0.6071,0.3023)
(0.6072,0.2805)
(0.6074,0.3218)
(0.6075,0.3190)
(0.6076,0.3369)
(0.6077,0.3380)
(0.6079,0.3013)
(0.6080,0.2650)
(0.6081,0.2604)
(0.6082,0.2822)
(0.6083,0.2699)
(0.6085,0.3255)
(0.6086,0.2887)
(0.6087,0.2776)
(0.6088,0.2900)
(0.6090,0.3176)
(0.6091,0.3222)
(0.6092,0.3236)
(0.6093,0.3032)
(0.6094,0.3177)
(0.6096,0.3139)
(0.6097,0.2882)
(0.6098,0.2984)
(0.6099,0.3345)
(0.6101,0.3242)
(0.6102,0.2815)
\PST@Solid(0.6102,0.2815)
(0.6103,0.2885)
(0.6104,0.3119)
(0.6105,0.3086)
(0.6107,0.3088)
(0.6108,0.3327)
(0.6109,0.2980)
(0.6110,0.3333)
(0.6111,0.2567)
(0.6112,0.3203)
(0.6114,0.3196)
(0.6115,0.2931)
(0.6116,0.3209)
(0.6117,0.2833)
(0.6118,0.2378)
(0.6120,0.2643)
(0.6121,0.3265)
(0.6122,0.2755)
(0.6123,0.2544)
(0.6124,0.3219)
(0.6125,0.2730)
(0.6127,0.3042)
(0.6128,0.3346)
(0.6129,0.3057)
(0.6130,0.3191)
(0.6131,0.3282)
(0.6132,0.3211)
(0.6133,0.2972)
(0.6135,0.2913)
(0.6136,0.2698)
(0.6137,0.2970)
(0.6138,0.2860)
(0.6139,0.3200)
(0.6140,0.2908)
(0.6141,0.3152)
(0.6143,0.2684)
(0.6144,0.3204)
(0.6145,0.3109)
(0.6146,0.2907)
(0.6147,0.2995)
(0.6148,0.3212)
(0.6149,0.3159)
(0.6151,0.3475)
(0.6152,0.2827)
(0.6153,0.3292)
(0.6154,0.3095)
(0.6155,0.3061)
(0.6156,0.3334)
(0.6157,0.2952)
(0.6158,0.2944)
(0.6159,0.2781)
(0.6161,0.2621)
(0.6162,0.2977)
(0.6163,0.3201)
(0.6164,0.3301)
(0.6165,0.2904)
(0.6166,0.3156)
(0.6167,0.3089)
(0.6168,0.3233)
(0.6169,0.3107)
(0.6170,0.3173)
(0.6172,0.2865)
(0.6173,0.2281)
(0.6174,0.2997)
(0.6175,0.3120)
(0.6176,0.2907)
(0.6177,0.2671)
(0.6178,0.3237)
(0.6179,0.2575)
(0.6180,0.3221)
(0.6181,0.3272)
(0.6182,0.3144)
(0.6183,0.2861)
(0.6185,0.2824)
(0.6186,0.3113)
(0.6187,0.3300)
(0.6188,0.2850)
(0.6189,0.2896)
(0.6190,0.2951)
(0.6191,0.3143)
(0.6192,0.3116)
(0.6193,0.3360)
(0.6194,0.2783)
(0.6195,0.2718)
(0.6196,0.3155)
(0.6197,0.3183)
(0.6198,0.2697)
(0.6199,0.2888)
(0.6200,0.3412)
(0.6201,0.3226)
(0.6202,0.3353)
(0.6204,0.2670)
(0.6205,0.3073)
(0.6206,0.3191)
(0.6207,0.3223)
(0.6208,0.2627)
(0.6209,0.3280)
(0.6210,0.3213)
(0.6211,0.3291)
(0.6212,0.3296)
(0.6213,0.3126)
\PST@Solid(0.6213,0.3126)
(0.6214,0.3197)
(0.6215,0.3356)
(0.6216,0.3092)
(0.6217,0.3181)
(0.6218,0.3329)
(0.6219,0.2936)
(0.6220,0.2843)
(0.6221,0.2506)
(0.6222,0.2943)
(0.6223,0.3217)
(0.6224,0.3186)
(0.6225,0.2739)
(0.6226,0.2263)
(0.6227,0.2094)
(0.6228,0.3178)
(0.6229,0.3184)
(0.6230,0.3276)
(0.6231,0.3082)
(0.6232,0.3232)
(0.6233,0.2751)
(0.6234,0.3151)
(0.6235,0.3211)
(0.6236,0.3157)
(0.6237,0.3183)
(0.6238,0.3289)
(0.6239,0.3397)
(0.6240,0.3140)
(0.6241,0.3282)
(0.6242,0.3044)
(0.6243,0.2948)
(0.6244,0.3412)
(0.6245,0.2979)
(0.6246,0.3123)
(0.6247,0.3219)
(0.6248,0.3314)
(0.6249,0.3293)
(0.6250,0.3330)
(0.6251,0.3238)
(0.6252,0.2858)
(0.6253,0.2898)
(0.6254,0.3080)
(0.6255,0.3109)
(0.6256,0.3198)
(0.6257,0.3302)
(0.6258,0.2689)
(0.6259,0.2598)
(0.6260,0.3327)
(0.6261,0.3064)
(0.6262,0.3115)
(0.6263,0.3254)
(0.6264,0.3286)
(0.6265,0.3435)
(0.6265,0.3047)
(0.6266,0.3159)
(0.6267,0.3316)
(0.6268,0.3443)
(0.6269,0.3247)
(0.6270,0.2891)
(0.6271,0.3490)
(0.6272,0.3125)
(0.6273,0.2949)
(0.6274,0.3333)
(0.6275,0.2640)
(0.6276,0.3471)
(0.6277,0.3431)
(0.6278,0.3196)
(0.6279,0.2995)
(0.6280,0.3373)
(0.6281,0.3456)
(0.6282,0.2866)
(0.6283,0.3136)
(0.6283,0.3220)
(0.6284,0.2553)
(0.6285,0.3519)
(0.6286,0.3108)
(0.6287,0.3418)
(0.6288,0.3067)
(0.6289,0.3504)
(0.6290,0.3373)
(0.6291,0.2907)
(0.6292,0.2785)
(0.6293,0.3150)
(0.6294,0.3171)
(0.6295,0.2989)
(0.6295,0.3284)
(0.6296,0.3363)
(0.6297,0.3196)
(0.6298,0.3401)
(0.6299,0.3450)
(0.6300,0.3040)
(0.6301,0.3381)
(0.6302,0.3065)
(0.6303,0.3237)
(0.6304,0.3394)
(0.6305,0.3074)
(0.6306,0.3092)
(0.6306,0.3090)
(0.6307,0.3043)
(0.6308,0.3194)
(0.6309,0.3489)
\PST@Solid(0.6309,0.3489)
(0.6310,0.3604)
(0.6311,0.3073)
(0.6312,0.2871)
(0.6313,0.3346)
(0.6314,0.2929)
(0.6315,0.3272)
(0.6315,0.3431)
(0.6316,0.3082)
(0.6317,0.3571)
(0.6318,0.3197)
(0.6319,0.3333)
(0.6320,0.3048)
(0.6321,0.3508)
(0.6322,0.3096)
(0.6323,0.2373)
(0.6323,0.3402)
(0.6324,0.3414)
(0.6325,0.3018)
(0.6326,0.3336)
(0.6327,0.3453)
(0.6328,0.3255)
(0.6329,0.2732)
(0.6330,0.3042)
(0.6330,0.3143)
(0.6331,0.3353)
(0.6332,0.2958)
(0.6333,0.3228)
(0.6334,0.3272)
(0.6335,0.3431)
(0.6336,0.3516)
(0.6337,0.3350)
(0.6337,0.2997)
(0.6338,0.3251)
(0.6339,0.2582)
(0.6340,0.2981)
(0.6341,0.3449)
(0.6342,0.3538)
(0.6343,0.3441)
(0.6343,0.3420)
(0.6344,0.3268)
(0.6345,0.3552)
(0.6346,0.2715)
(0.6347,0.3216)
(0.6348,0.3123)
(0.6349,0.3146)
(0.6349,0.3090)
(0.6350,0.3432)
(0.6351,0.3335)
(0.6352,0.2850)
(0.6353,0.2681)
(0.6354,0.2984)
(0.6355,0.3389)
(0.6355,0.3529)
(0.6356,0.3637)
(0.6357,0.3667)
(0.6358,0.3096)
(0.6359,0.3081)
(0.6360,0.3556)
(0.6360,0.3600)
(0.6361,0.2967)
(0.6362,0.3155)
(0.6363,0.3585)
(0.6364,0.3464)
(0.6365,0.3469)
(0.6365,0.3518)
(0.6366,0.3532)
(0.6367,0.3152)
(0.6368,0.3408)
(0.6369,0.3422)
(0.6370,0.3602)
(0.6370,0.3142)
(0.6371,0.3422)
(0.6372,0.3524)
(0.6373,0.3434)
(0.6374,0.3611)
(0.6375,0.3502)
(0.6375,0.2894)
(0.6376,0.3553)
(0.6377,0.3413)
(0.6378,0.3574)
(0.6379,0.3335)
(0.6379,0.3195)
(0.6380,0.3434)
(0.6381,0.3707)
(0.6382,0.3445)
(0.6383,0.3403)
(0.6384,0.3523)
(0.6384,0.3357)
(0.6385,0.3600)
(0.6386,0.3502)
(0.6387,0.2810)
(0.6388,0.3542)
(0.6388,0.3521)
(0.6389,0.3100)
(0.6390,0.3534)
(0.6391,0.3571)
(0.6392,0.3423)
(0.6392,0.3429)
(0.6393,0.3461)
(0.6394,0.3574)
\PST@Solid(0.6394,0.3574)
(0.6395,0.3222)
(0.6396,0.3066)
(0.6396,0.3089)
(0.6397,0.3616)
(0.6398,0.3517)
(0.6399,0.2832)
(0.6400,0.3587)
(0.6400,0.3781)
(0.6401,0.3543)
(0.6402,0.3844)
(0.6403,0.3380)
(0.6404,0.3482)
(0.6404,0.3701)
(0.6405,0.3807)
(0.6406,0.3675)
(0.6407,0.3680)
(0.6408,0.3510)
(0.6408,0.3767)
(0.6409,0.2998)
(0.6410,0.3428)
(0.6411,0.3529)
(0.6411,0.3578)
(0.6412,0.3712)
(0.6413,0.3587)
(0.6414,0.2848)
(0.6415,0.3748)
(0.6415,0.3612)
(0.6416,0.3660)
(0.6417,0.3843)
(0.6418,0.3494)
(0.6418,0.3716)
(0.6419,0.3724)
(0.6420,0.3257)
(0.6421,0.3544)
(0.6422,0.3193)
(0.6422,0.3465)
(0.6423,0.3295)
(0.6424,0.3914)
(0.6425,0.3722)
(0.6425,0.3819)
(0.6426,0.3898)
(0.6427,0.3804)
(0.6428,0.3591)
(0.6428,0.3547)
(0.6429,0.3512)
(0.6430,0.4026)
(0.6431,0.3635)
(0.6432,0.3559)
(0.6432,0.3646)
(0.6433,0.3702)
(0.6434,0.4008)
(0.6435,0.3673)
(0.6435,0.3659)
(0.6436,0.3798)
(0.6437,0.3654)
(0.6438,0.4192)
(0.6438,0.4096)
(0.6439,0.3848)
(0.6440,0.3887)
(0.6441,0.3461)
(0.6441,0.4061)
(0.6442,0.4175)
(0.6443,0.4029)
(0.6444,0.4062)
(0.6444,0.3771)
(0.6445,0.3727)
(0.6446,0.3810)
(0.6447,0.4092)
(0.6447,0.3973)
(0.6448,0.4074)
(0.6449,0.3775)
(0.6450,0.3231)
(0.6450,0.3893)
(0.6451,0.4134)
(0.6452,0.4254)
(0.6452,0.3929)
(0.6453,0.4059)
(0.6454,0.4063)
(0.6455,0.4046)
(0.6455,0.4106)
(0.6456,0.4417)
(0.6457,0.4323)
(0.6458,0.3162)
(0.6458,0.4459)
(0.6459,0.4355)
(0.6460,0.4295)
(0.6461,0.4249)
(0.6461,0.4448)
(0.6462,0.3966)
(0.6463,0.4697)
(0.6463,0.4432)
(0.6464,0.4476)
(0.6465,0.4538)
(0.6466,0.4415)
(0.6466,0.4774)
(0.6467,0.4807)
(0.6468,0.4541)
(0.6469,0.5132)
(0.6469,0.5253)

\rput[r](0.5200,0.4850){$3000x^{-1}$}
\PST@Dashed(0.5360,0.4850)
(0.6150,0.4850)

\PST@Dashed(0.1490,0.5218)
(0.1490,0.5218)
(0.1540,0.5191)
(0.1591,0.5164)
(0.1641,0.5137)
(0.1691,0.5111)
(0.1741,0.5084)
(0.1792,0.5057)
(0.1842,0.5030)
(0.1892,0.5004)
(0.1943,0.4977)
(0.1993,0.4950)
(0.2043,0.4923)
(0.2094,0.4896)
(0.2144,0.4870)
(0.2194,0.4843)
(0.2244,0.4816)
(0.2295,0.4789)
(0.2345,0.4762)
(0.2395,0.4736)
(0.2446,0.4709)
(0.2496,0.4682)
(0.2546,0.4655)
(0.2597,0.4629)
(0.2647,0.4602)
(0.2697,0.4575)
(0.2747,0.4548)
(0.2798,0.4521)
(0.2848,0.4495)
(0.2898,0.4468)
(0.2949,0.4441)
(0.2999,0.4414)
(0.3049,0.4387)
(0.3099,0.4361)
(0.3150,0.4334)
(0.3200,0.4307)
(0.3250,0.4280)
(0.3301,0.4254)
(0.3351,0.4227)
(0.3401,0.4200)
(0.3452,0.4173)
(0.3502,0.4146)
(0.3552,0.4120)
(0.3602,0.4093)
(0.3653,0.4066)
(0.3703,0.4039)
(0.3753,0.4012)
(0.3804,0.3986)
(0.3854,0.3959)
(0.3904,0.3932)
(0.3954,0.3905)
(0.4005,0.3879)
(0.4055,0.3852)
(0.4105,0.3825)
(0.4156,0.3798)
(0.4206,0.3771)
(0.4256,0.3745)
(0.4307,0.3718)
(0.4357,0.3691)
(0.4407,0.3664)
(0.4457,0.3638)
(0.4508,0.3611)
(0.4558,0.3584)
(0.4608,0.3557)
(0.4659,0.3530)
(0.4709,0.3504)
(0.4759,0.3477)
(0.4810,0.3450)
(0.4860,0.3423)
(0.4910,0.3396)
(0.4960,0.3370)
(0.5011,0.3343)
(0.5061,0.3316)
(0.5111,0.3289)
(0.5162,0.3263)
(0.5212,0.3236)
(0.5262,0.3209)
(0.5312,0.3182)
(0.5363,0.3155)
(0.5413,0.3129)
(0.5463,0.3102)
(0.5514,0.3075)
(0.5564,0.3048)
(0.5614,0.3021)
(0.5665,0.2995)
(0.5715,0.2968)
(0.5765,0.2941)
(0.5815,0.2914)
(0.5866,0.2888)
(0.5916,0.2861)
(0.5966,0.2834)
(0.6017,0.2807)
(0.6067,0.2780)
(0.6117,0.2754)
(0.6168,0.2727)
(0.6218,0.2700)
(0.6268,0.2673)
(0.6318,0.2647)
(0.6369,0.2620)
(0.6419,0.2593)
(0.6469,0.2566)

\catcode`@=12
\fi
\endpspicture

%% file: wwhisto.tex
\ifx\PSTloaded\undefined
\def\PSTloaded{t}
\psset{arrowsize=.01 3.2 1.4 .3}
\psset{dotsize=.01}
\catcode`@=11

\newpsobject{PST@Border}{psline}{linewidth=.0015,linestyle=solid}
\newpsobject{PST@Axes}{psline}{linewidth=.0015,linestyle=dotted,dotsep=.004}
\newpsobject{PST@Solid}{psline}{linewidth=.0015,linestyle=solid}
\newpsobject{PST@Dashed}{psline}{linewidth=.0015,linestyle=dashed,dash=.01 .01}
\newpsobject{PST@Dotted}{psline}{linewidth=.0025,linestyle=dotted,dotsep=.008}
\newpsobject{PST@LongDash}{psline}{linewidth=.0015,linestyle=dashed,dash=.02 .01}
\newpsobject{PST@Diamond}{psdots}{linewidth=.001,linestyle=solid,dotstyle=square,dotangle=45}
\newpsobject{PST@Filldiamond}{psdots}{linewidth=.001,linestyle=solid,dotstyle=square*,dotangle=45}
\newpsobject{PST@Cross}{psdots}{linewidth=.001,linestyle=solid,dotstyle=+,dotangle=45}
\newpsobject{PST@Plus}{psdots}{linewidth=.001,linestyle=solid,dotstyle=+}
\newpsobject{PST@Square}{psdots}{linewidth=.001,linestyle=solid,dotstyle=square}
\newpsobject{PST@Circle}{psdots}{linewidth=.001,linestyle=solid,dotstyle=o}
\newpsobject{PST@Triangle}{psdots}{linewidth=.001,linestyle=solid,dotstyle=triangle}
\newpsobject{PST@Pentagon}{psdots}{linewidth=.001,linestyle=solid,dotstyle=pentagon}
\newpsobject{PST@Fillsquare}{psdots}{linewidth=.001,linestyle=solid,dotstyle=square*}
\newpsobject{PST@Fillcircle}{psdots}{linewidth=.001,linestyle=solid,dotstyle=*}
\newpsobject{PST@Filltriangle}{psdots}{linewidth=.001,linestyle=solid,dotstyle=triangle*}
\newpsobject{PST@Fillpentagon}{psdots}{linewidth=.001,linestyle=solid,dotstyle=pentagon*}
\newpsobject{PST@Arrow}{psline}{linewidth=.001,linestyle=solid}
\catcode`@=12

\fi
\psset{unit=5.0in,xunit=5.0in,yunit=3.0in}
\pspicture(0.000000,0.000000)(0.700000,0.700000)
\ifx\nofigs\undefined
\catcode`@=11

\PST@Border(0.1490,0.1260)
(0.1640,0.1260)

\PST@Border(0.6470,0.1260)
(0.6320,0.1260)

\rput[r](0.1330,0.1260){ 1}
\PST@Border(0.1490,0.1668)
(0.1565,0.1668)

\PST@Border(0.6470,0.1668)
(0.6395,0.1668)

\PST@Border(0.1490,0.1906)
(0.1565,0.1906)

\PST@Border(0.6470,0.1906)
(0.6395,0.1906)

\PST@Border(0.1490,0.2076)
(0.1565,0.2076)

\PST@Border(0.6470,0.2076)
(0.6395,0.2076)

\PST@Border(0.1490,0.2207)
(0.1565,0.2207)

\PST@Border(0.6470,0.2207)
(0.6395,0.2207)

\PST@Border(0.1490,0.2314)
(0.1565,0.2314)

\PST@Border(0.6470,0.2314)
(0.6395,0.2314)

\PST@Border(0.1490,0.2405)
(0.1565,0.2405)

\PST@Border(0.6470,0.2405)
(0.6395,0.2405)

\PST@Border(0.1490,0.2484)
(0.1565,0.2484)

\PST@Border(0.6470,0.2484)
(0.6395,0.2484)

\PST@Border(0.1490,0.2553)
(0.1565,0.2553)

\PST@Border(0.6470,0.2553)
(0.6395,0.2553)

\PST@Border(0.1490,0.2615)
(0.1640,0.2615)

\PST@Border(0.6470,0.2615)
(0.6320,0.2615)

\rput[r](0.1330,0.2615){ 10}
\PST@Border(0.1490,0.3023)
(0.1565,0.3023)

\PST@Border(0.6470,0.3023)
(0.6395,0.3023)

\PST@Border(0.1490,0.3261)
(0.1565,0.3261)

\PST@Border(0.6470,0.3261)
(0.6395,0.3261)

\PST@Border(0.1490,0.3431)
(0.1565,0.3431)

\PST@Border(0.6470,0.3431)
(0.6395,0.3431)

\PST@Border(0.1490,0.3562)
(0.1565,0.3562)

\PST@Border(0.6470,0.3562)
(0.6395,0.3562)

\PST@Border(0.1490,0.3669)
(0.1565,0.3669)

\PST@Border(0.6470,0.3669)
(0.6395,0.3669)

\PST@Border(0.1490,0.3760)
(0.1565,0.3760)

\PST@Border(0.6470,0.3760)
(0.6395,0.3760)

\PST@Border(0.1490,0.3839)
(0.1565,0.3839)

\PST@Border(0.6470,0.3839)
(0.6395,0.3839)

\PST@Border(0.1490,0.3908)
(0.1565,0.3908)

\PST@Border(0.6470,0.3908)
(0.6395,0.3908)

\PST@Border(0.1490,0.3970)
(0.1640,0.3970)

\PST@Border(0.6470,0.3970)
(0.6320,0.3970)

\rput[r](0.1330,0.3970){ 100}
\PST@Border(0.1490,0.4378)
(0.1565,0.4378)

\PST@Border(0.6470,0.4378)
(0.6395,0.4378)

\PST@Border(0.1490,0.4616)
(0.1565,0.4616)

\PST@Border(0.6470,0.4616)
(0.6395,0.4616)

\PST@Border(0.1490,0.4786)
(0.1565,0.4786)

\PST@Border(0.6470,0.4786)
(0.6395,0.4786)

\PST@Border(0.1490,0.4917)
(0.1565,0.4917)

\PST@Border(0.6470,0.4917)
(0.6395,0.4917)

\PST@Border(0.1490,0.5024)
(0.1565,0.5024)

\PST@Border(0.6470,0.5024)
(0.6395,0.5024)

\PST@Border(0.1490,0.5115)
(0.1565,0.5115)

\PST@Border(0.6470,0.5115)
(0.6395,0.5115)

\PST@Border(0.1490,0.5194)
(0.1565,0.5194)

\PST@Border(0.6470,0.5194)
(0.6395,0.5194)

\PST@Border(0.1490,0.5263)
(0.1565,0.5263)

\PST@Border(0.6470,0.5263)
(0.6395,0.5263)

\PST@Border(0.1490,0.5325)
(0.1640,0.5325)

\PST@Border(0.6470,0.5325)
(0.6320,0.5325)

\rput[r](0.1330,0.5325){ 1000}
\PST@Border(0.1490,0.5733)
(0.1565,0.5733)

\PST@Border(0.6470,0.5733)
(0.6395,0.5733)

\PST@Border(0.1490,0.5971)
(0.1565,0.5971)

\PST@Border(0.6470,0.5971)
(0.6395,0.5971)

\PST@Border(0.1490,0.6141)
(0.1565,0.6141)

\PST@Border(0.6470,0.6141)
(0.6395,0.6141)

\PST@Border(0.1490,0.6272)
(0.1565,0.6272)

\PST@Border(0.6470,0.6272)
(0.6395,0.6272)

\PST@Border(0.1490,0.6379)
(0.1565,0.6379)

\PST@Border(0.6470,0.6379)
(0.6395,0.6379)

\PST@Border(0.1490,0.6470)
(0.1565,0.6470)

\PST@Border(0.6470,0.6470)
(0.6395,0.6470)

\PST@Border(0.1490,0.6549)
(0.1565,0.6549)

\PST@Border(0.6470,0.6549)
(0.6395,0.6549)

\PST@Border(0.1490,0.6618)
(0.1565,0.6618)

\PST@Border(0.6470,0.6618)
(0.6395,0.6618)

\PST@Border(0.1490,0.6680)
(0.1640,0.6680)

\PST@Border(0.6470,0.6680)
(0.6320,0.6680)

\rput[r](0.1330,0.6680){ 10000}
\PST@Border(0.1490,0.1260)
(0.1490,0.1460)

\PST@Border(0.1490,0.6680)
(0.1490,0.6480)

\rput(0.1490,0.0840){ 1}
\PST@Border(0.1740,0.1260)
(0.1740,0.1360)

\PST@Border(0.1740,0.6680)
(0.1740,0.6580)

\PST@Border(0.2070,0.1260)
(0.2070,0.1360)

\PST@Border(0.2070,0.6680)
(0.2070,0.6580)

\PST@Border(0.2240,0.1260)
(0.2240,0.1360)

\PST@Border(0.2240,0.6680)
(0.2240,0.6580)

\PST@Border(0.2320,0.1260)
(0.2320,0.1460)

\PST@Border(0.2320,0.6680)
(0.2320,0.6480)

\rput(0.2320,0.0840){ 10}
\PST@Border(0.2570,0.1260)
(0.2570,0.1360)

\PST@Border(0.2570,0.6680)
(0.2570,0.6580)

\PST@Border(0.2900,0.1260)
(0.2900,0.1360)

\PST@Border(0.2900,0.6680)
(0.2900,0.6580)

\PST@Border(0.3070,0.1260)
(0.3070,0.1360)

\PST@Border(0.3070,0.6680)
(0.3070,0.6580)

\PST@Border(0.3150,0.1260)
(0.3150,0.1460)

\PST@Border(0.3150,0.6680)
(0.3150,0.6480)

\rput(0.3150,0.0840){ 100}
\PST@Border(0.3400,0.1260)
(0.3400,0.1360)

\PST@Border(0.3400,0.6680)
(0.3400,0.6580)

\PST@Border(0.3730,0.1260)
(0.3730,0.1360)

\PST@Border(0.3730,0.6680)
(0.3730,0.6580)

\PST@Border(0.3900,0.1260)
(0.3900,0.1360)

\PST@Border(0.3900,0.6680)
(0.3900,0.6580)

\PST@Border(0.3980,0.1260)
(0.3980,0.1460)

\PST@Border(0.3980,0.6680)
(0.3980,0.6480)

\rput(0.3980,0.0840){ 1000}
\PST@Border(0.4230,0.1260)
(0.4230,0.1360)

\PST@Border(0.4230,0.6680)
(0.4230,0.6580)

\PST@Border(0.4560,0.1260)
(0.4560,0.1360)

\PST@Border(0.4560,0.6680)
(0.4560,0.6580)

\PST@Border(0.4730,0.1260)
(0.4730,0.1360)

\PST@Border(0.4730,0.6680)
(0.4730,0.6580)

\PST@Border(0.4810,0.1260)
(0.4810,0.1460)

\PST@Border(0.4810,0.6680)
(0.4810,0.6480)

\rput(0.4810,0.0840){ 10000}
\PST@Border(0.5060,0.1260)
(0.5060,0.1360)

\PST@Border(0.5060,0.6680)
(0.5060,0.6580)

\PST@Border(0.5390,0.1260)
(0.5390,0.1360)

\PST@Border(0.5390,0.6680)
(0.5390,0.6580)

\PST@Border(0.5560,0.1260)
(0.5560,0.1360)

\PST@Border(0.5560,0.6680)
(0.5560,0.6580)

\PST@Border(0.5640,0.1260)
(0.5640,0.1460)

\PST@Border(0.5640,0.6680)
(0.5640,0.6480)

\rput(0.5640,0.0840){ 100000}
\PST@Border(0.5890,0.1260)
(0.5890,0.1360)

\PST@Border(0.5890,0.6680)
(0.5890,0.6580)

\PST@Border(0.6220,0.1260)
(0.6220,0.1360)

\PST@Border(0.6220,0.6680)
(0.6220,0.6580)

\PST@Border(0.6390,0.1260)
(0.6390,0.1360)

\PST@Border(0.6390,0.6680)
(0.6390,0.6580)

\PST@Border(0.6470,0.1260)
(0.6470,0.1460)

\PST@Border(0.6470,0.6680)
(0.6470,0.6480)

\rput(0.6470,0.0840){$10^{6}$}
\PST@Border(0.1490,0.1260)
(0.6470,0.1260)
(0.6470,0.6680)
(0.1490,0.6680)
(0.1490,0.1260)

\rput(0.3980,0.0210){$\tau$}
\rput[r](0.6310,0.6270){$c=0.2$}
\PST@Diamond(0.1490,0.4772)
\PST@Diamond(0.1733,0.4862)
\PST@Diamond(0.1872,0.4780)
\PST@Diamond(0.1976,0.4753)
\PST@Diamond(0.2045,0.4753)
\PST@Diamond(0.2115,0.4786)
\PST@Diamond(0.2184,0.4842)
\PST@Diamond(0.2219,0.4838)
\PST@Diamond(0.2254,0.4716)
\PST@Diamond(0.2288,0.4777)
\PST@Diamond(0.2323,0.4662)
\PST@Diamond(0.2358,0.4660)
\PST@Diamond(0.2393,0.4714)
\PST@Diamond(0.2427,0.4763)
\PST@Diamond(0.2462,0.5042)
\PST@Diamond(0.2497,0.4641)
\PST@Diamond(0.2531,0.5009)
\PST@Diamond(0.2566,0.5048)
\PST@Diamond(0.2601,0.5272)
\PST@Diamond(0.2636,0.4904)
\PST@Diamond(0.2670,0.5129)
\PST@Diamond(0.2705,0.5095)
\PST@Diamond(0.2740,0.5091)
\PST@Diamond(0.2774,0.5056)
\PST@Diamond(0.2809,0.5191)
\PST@Diamond(0.2844,0.5316)
\PST@Diamond(0.2879,0.5142)
\PST@Diamond(0.2913,0.5359)
\PST@Diamond(0.2948,0.5234)
\PST@Diamond(0.2983,0.5374)
\PST@Diamond(0.3017,0.5369)
\PST@Diamond(0.3052,0.5320)
\PST@Diamond(0.3087,0.5447)
\PST@Diamond(0.3122,0.5404)
\PST@Diamond(0.3156,0.5519)
\PST@Diamond(0.3191,0.5461)
\PST@Diamond(0.3226,0.5486)
\PST@Diamond(0.3260,0.5564)
\PST@Diamond(0.3295,0.5556)
\PST@Diamond(0.3330,0.5547)
\PST@Diamond(0.3365,0.5589)
\PST@Diamond(0.3399,0.5566)
\PST@Diamond(0.3434,0.5651)
\PST@Diamond(0.3469,0.5639)
\PST@Diamond(0.3503,0.5677)
\PST@Diamond(0.3538,0.5663)
\PST@Diamond(0.3573,0.5683)
\PST@Diamond(0.3608,0.5668)
\PST@Diamond(0.3642,0.5718)
\PST@Diamond(0.3677,0.5683)
\PST@Diamond(0.3712,0.5701)
\PST@Diamond(0.3746,0.5713)
\PST@Diamond(0.3781,0.5685)
\PST@Diamond(0.3816,0.5661)
\PST@Diamond(0.3851,0.5667)
\PST@Diamond(0.3885,0.5646)
\PST@Diamond(0.3920,0.5632)
\PST@Diamond(0.3955,0.5608)
\PST@Diamond(0.3990,0.5557)
\PST@Diamond(0.4024,0.5507)
\PST@Diamond(0.4059,0.5505)
\PST@Diamond(0.4094,0.5439)
\PST@Diamond(0.4128,0.5394)
\PST@Diamond(0.4163,0.5293)
\PST@Diamond(0.4198,0.5242)
\PST@Diamond(0.4233,0.5114)
\PST@Diamond(0.4267,0.5027)
\PST@Diamond(0.4302,0.4974)
\PST@Diamond(0.4337,0.4867)
\PST@Diamond(0.4371,0.4658)
\PST@Diamond(0.4406,0.4601)
\PST@Diamond(0.4441,0.4492)
\PST@Diamond(0.4476,0.4326)
\PST@Diamond(0.4510,0.4082)
\PST@Diamond(0.4545,0.3846)
\PST@Diamond(0.4580,0.3513)
\PST@Diamond(0.4614,0.3352)
\PST@Diamond(0.4649,0.3242)
\PST@Diamond(0.4684,0.2813)
\PST@Diamond(0.4719,0.2813)
\PST@Diamond(0.4753,0.1906)
\PST@Diamond(0.4788,0.1668)
\PST@Diamond(0.4823,0.1668)
\PST@Diamond(0.4857,0.1668)
\PST@Diamond(0.4892,0.1906)
\PST@Diamond(0.5595,0.6270)
\rput[r](0.6310,0.5850){$c=0.4$}
\PST@Plus(0.1490,0.5166)
\PST@Plus(0.1718,0.5242)
\PST@Plus(0.1870,0.5233)
\PST@Plus(0.1984,0.5248)
\PST@Plus(0.2060,0.5207)
\PST@Plus(0.2098,0.5273)
\PST@Plus(0.2174,0.5265)
\PST@Plus(0.2212,0.5262)
\PST@Plus(0.2250,0.5287)
\PST@Plus(0.2288,0.5255)
\PST@Plus(0.2326,0.5252)
\PST@Plus(0.2364,0.5260)
\PST@Plus(0.2402,0.5223)
\PST@Plus(0.2441,0.5637)
\PST@Plus(0.2479,0.5611)
\PST@Plus(0.2517,0.5606)
\PST@Plus(0.2555,0.5614)
\PST@Plus(0.2593,0.5571)
\PST@Plus(0.2631,0.5778)
\PST@Plus(0.2669,0.5759)
\PST@Plus(0.2707,0.5765)
\PST@Plus(0.2745,0.5901)
\PST@Plus(0.2783,0.5876)
\PST@Plus(0.2821,0.5839)
\PST@Plus(0.2859,0.5980)
\PST@Plus(0.2897,0.6020)
\PST@Plus(0.2935,0.5980)
\PST@Plus(0.2973,0.5958)
\PST@Plus(0.3011,0.6106)
\PST@Plus(0.3049,0.6094)
\PST@Plus(0.3087,0.6165)
\PST@Plus(0.3125,0.6135)
\PST@Plus(0.3163,0.6191)
\PST@Plus(0.3201,0.6186)
\PST@Plus(0.3239,0.6268)
\PST@Plus(0.3277,0.6276)
\PST@Plus(0.3315,0.6265)
\PST@Plus(0.3353,0.6311)
\PST@Plus(0.3391,0.6311)
\PST@Plus(0.3429,0.6347)
\PST@Plus(0.3467,0.6365)
\PST@Plus(0.3505,0.6377)
\PST@Plus(0.3543,0.6379)
\PST@Plus(0.3581,0.6398)
\PST@Plus(0.3619,0.6413)
\PST@Plus(0.3657,0.6398)
\PST@Plus(0.3695,0.6410)
\PST@Plus(0.3733,0.6396)
\PST@Plus(0.3771,0.6384)
\PST@Plus(0.3809,0.6388)
\PST@Plus(0.3847,0.6374)
\PST@Plus(0.3885,0.6363)
\PST@Plus(0.3923,0.6319)
\PST@Plus(0.3961,0.6269)
\PST@Plus(0.3999,0.6238)
\PST@Plus(0.4037,0.6198)
\PST@Plus(0.4075,0.6117)
\PST@Plus(0.4113,0.6063)
\PST@Plus(0.4151,0.5994)
\PST@Plus(0.4189,0.5893)
\PST@Plus(0.4227,0.5799)
\PST@Plus(0.4265,0.5678)
\PST@Plus(0.4304,0.5595)
\PST@Plus(0.4342,0.5459)
\PST@Plus(0.4380,0.5310)
\PST@Plus(0.4418,0.5138)
\PST@Plus(0.4456,0.4954)
\PST@Plus(0.4494,0.4790)
\PST@Plus(0.4532,0.4592)
\PST@Plus(0.4570,0.4390)
\PST@Plus(0.4608,0.4164)
\PST@Plus(0.4646,0.3999)
\PST@Plus(0.4684,0.3679)
\PST@Plus(0.4722,0.3460)
\PST@Plus(0.4760,0.3105)
\PST@Plus(0.4798,0.2892)
\PST@Plus(0.4836,0.2722)
\PST@Plus(0.4874,0.2671)
\PST@Plus(0.4912,0.2207)
\PST@Plus(0.4950,0.1906)
\PST@Plus(0.4988,0.1906)
\PST@Plus(0.5026,0.1260)
\PST@Plus(0.5064,0.1668)
\PST@Plus(0.5102,0.1260)
\PST@Plus(0.5140,0.1260)
\PST@Plus(0.5216,0.1260)
\PST@Plus(0.5595,0.5850)
\rput[r](0.6310,0.5430){$c=0.6$}
\PST@Square(0.1490,0.5036)
\PST@Square(0.1728,0.5108)
\PST@Square(0.1886,0.5061)
\PST@Square(0.1965,0.5033)
\PST@Square(0.2044,0.5063)
\PST@Square(0.2124,0.5059)
\PST@Square(0.2163,0.4988)
\PST@Square(0.2203,0.5060)
\PST@Square(0.2282,0.5482)
\PST@Square(0.2322,0.5091)
\PST@Square(0.2361,0.5057)
\PST@Square(0.2401,0.5000)
\PST@Square(0.2440,0.5459)
\PST@Square(0.2480,0.5438)
\PST@Square(0.2520,0.5414)
\PST@Square(0.2559,0.5423)
\PST@Square(0.2599,0.5660)
\PST@Square(0.2638,0.5355)
\PST@Square(0.2678,0.5801)
\PST@Square(0.2718,0.5581)
\PST@Square(0.2757,0.5734)
\PST@Square(0.2797,0.5716)
\PST@Square(0.2836,0.5858)
\PST@Square(0.2876,0.5908)
\PST@Square(0.2916,0.5889)
\PST@Square(0.2955,0.5951)
\PST@Square(0.2995,0.5914)
\PST@Square(0.3034,0.5971)
\PST@Square(0.3074,0.6064)
\PST@Square(0.3114,0.6034)
\PST@Square(0.3153,0.6132)
\PST@Square(0.3193,0.6121)
\PST@Square(0.3232,0.6164)
\PST@Square(0.3272,0.6166)
\PST@Square(0.3312,0.6186)
\PST@Square(0.3351,0.6270)
\PST@Square(0.3391,0.6231)
\PST@Square(0.3430,0.6284)
\PST@Square(0.3470,0.6309)
\PST@Square(0.3510,0.6306)
\PST@Square(0.3549,0.6318)
\PST@Square(0.3589,0.6350)
\PST@Square(0.3628,0.6361)
\PST@Square(0.3668,0.6370)
\PST@Square(0.3708,0.6391)
\PST@Square(0.3747,0.6412)
\PST@Square(0.3787,0.6410)
\PST@Square(0.3826,0.6392)
\PST@Square(0.3866,0.6394)
\PST@Square(0.3906,0.6406)
\PST@Square(0.3945,0.6396)
\PST@Square(0.3985,0.6378)
\PST@Square(0.4024,0.6343)
\PST@Square(0.4064,0.6330)
\PST@Square(0.4104,0.6288)
\PST@Square(0.4143,0.6259)
\PST@Square(0.4183,0.6207)
\PST@Square(0.4222,0.6167)
\PST@Square(0.4262,0.6106)
\PST@Square(0.4302,0.6016)
\PST@Square(0.4341,0.5942)
\PST@Square(0.4381,0.5876)
\PST@Square(0.4420,0.5785)
\PST@Square(0.4460,0.5663)
\PST@Square(0.4500,0.5528)
\PST@Square(0.4539,0.5383)
\PST@Square(0.4579,0.5295)
\PST@Square(0.4618,0.5154)
\PST@Square(0.4658,0.4954)
\PST@Square(0.4698,0.4792)
\PST@Square(0.4737,0.4537)
\PST@Square(0.4777,0.4415)
\PST@Square(0.4816,0.4213)
\PST@Square(0.4856,0.4015)
\PST@Square(0.4896,0.3793)
\PST@Square(0.4935,0.3416)
\PST@Square(0.4975,0.3318)
\PST@Square(0.5014,0.3199)
\PST@Square(0.5054,0.2671)
\PST@Square(0.5094,0.2484)
\PST@Square(0.5133,0.1906)
\PST@Square(0.5173,0.1668)
\PST@Square(0.5212,0.2076)
\PST@Square(0.5252,0.1260)
\PST@Square(0.5371,0.1668)
\PST@Square(0.5595,0.5430)
\rput[r](0.6310,0.5010){$c=0.8$}
\PST@Cross(0.1734,0.6519)
\PST@Cross(0.1881,0.6450)
\PST@Cross(0.1978,0.6373)
\PST@Cross(0.2027,0.6265)
\PST@Cross(0.2125,0.6155)
\PST@Cross(0.2173,0.6061)
\PST@Cross(0.2222,0.5982)
\PST@Cross(0.2271,0.5877)
\PST@Cross(0.2320,0.6198)
\PST@Cross(0.2369,0.6107)
\PST@Cross(0.2418,0.6047)
\PST@Cross(0.2466,0.5951)
\PST@Cross(0.2515,0.5925)
\PST@Cross(0.2564,0.6115)
\PST@Cross(0.2613,0.6086)
\PST@Cross(0.2662,0.6205)
\PST@Cross(0.2710,0.6161)
\PST@Cross(0.2759,0.6261)
\PST@Cross(0.2808,0.6336)
\PST@Cross(0.2857,0.6288)
\PST@Cross(0.2906,0.6398)
\PST@Cross(0.2955,0.6362)
\PST@Cross(0.3003,0.6436)
\PST@Cross(0.3052,0.6460)
\PST@Cross(0.3101,0.6443)
\PST@Cross(0.3150,0.6528)
\PST@Cross(0.3199,0.6514)
\PST@Cross(0.3247,0.6528)
\PST@Cross(0.3296,0.6520)
\PST@Cross(0.3345,0.6549)
\PST@Cross(0.3394,0.6558)
\PST@Cross(0.3443,0.6559)
\PST@Cross(0.3492,0.6593)
\PST@Cross(0.3540,0.6581)
\PST@Cross(0.3589,0.6551)
\PST@Cross(0.3638,0.6549)
\PST@Cross(0.3687,0.6545)
\PST@Cross(0.3736,0.6532)
\PST@Cross(0.3784,0.6524)
\PST@Cross(0.3833,0.6485)
\PST@Cross(0.3882,0.6469)
\PST@Cross(0.3931,0.6439)
\PST@Cross(0.3980,0.6414)
\PST@Cross(0.4029,0.6392)
\PST@Cross(0.4077,0.6364)
\PST@Cross(0.4126,0.6313)
\PST@Cross(0.4175,0.6302)
\PST@Cross(0.4224,0.6260)
\PST@Cross(0.4273,0.6215)
\PST@Cross(0.4321,0.6176)
\PST@Cross(0.4370,0.6130)
\PST@Cross(0.4419,0.6065)
\PST@Cross(0.4468,0.6007)
\PST@Cross(0.4517,0.5944)
\PST@Cross(0.4566,0.5883)
\PST@Cross(0.4614,0.5822)
\PST@Cross(0.4663,0.5760)
\PST@Cross(0.4712,0.5645)
\PST@Cross(0.4761,0.5583)
\PST@Cross(0.4810,0.5468)
\PST@Cross(0.4858,0.5357)
\PST@Cross(0.4907,0.5268)
\PST@Cross(0.4956,0.5153)
\PST@Cross(0.5005,0.5077)
\PST@Cross(0.5054,0.4908)
\PST@Cross(0.5103,0.4771)
\PST@Cross(0.5151,0.4632)
\PST@Cross(0.5200,0.4478)
\PST@Cross(0.5249,0.4372)
\PST@Cross(0.5298,0.4289)
\PST@Cross(0.5347,0.4224)
\PST@Cross(0.5396,0.3958)
\PST@Cross(0.5444,0.3777)
\PST@Cross(0.5493,0.3550)
\PST@Cross(0.5542,0.3473)
\PST@Cross(0.5591,0.3385)
\PST@Cross(0.5640,0.3199)
\PST@Cross(0.5688,0.3052)
\PST@Cross(0.5737,0.3079)
\PST@Cross(0.5786,0.2854)
\PST@Cross(0.5835,0.2484)
\PST@Cross(0.5884,0.2314)
\PST@Cross(0.5933,0.2076)
\PST@Cross(0.5981,0.2076)
\PST@Cross(0.6030,0.1260)
\PST@Cross(0.6079,0.1260)
\PST@Cross(0.6274,0.1260)
\PST@Cross(0.5595,0.5010)
\catcode`@=12
\fi
\endpspicture